\documentclass[12pt,a4paper]{article}
\usepackage{jheppub}
\pdfoutput=1
\usepackage{dsfont}
\usepackage{amssymb,amsmath}
\usepackage{epsfig}
\usepackage{epstopdf}
\usepackage{latexsym}
\usepackage{graphicx}
\usepackage{subfigure}
\usepackage{booktabs}
\usepackage{bbm}
\makeatletter
\def\@fpheader{\relax}
\makeatother

\newcommand{\lP}{\ensuremath{l_P}} 

\newcommand{\cO}{{\mathcal{O}}}

\newcommand{\be}{\begin{equation}}
\newcommand{\ee}{\end{equation}}
\newcommand{\bea}{\begin{eqnarray}}
\newcommand{\eea}{\end{eqnarray}}

\subheader{\begin{flushright}
UTTG-14-15\\
TCC-005-15
\end{flushright}}

\title{Holographic entanglement entropy and the extended phase structure of STU black holes}
\author[a,b]{Elena Caceres,}
\author[b,c]{Phuc H. Nguyen}
\author[b,c]{and Juan F. Pedraza}
\affiliation[a]{Facultad de Ciencias, Universidad de Colima, Bernal Diaz del Castillo 340, Colima, Mexico}
\affiliation[b]{Theory Group, Department of Physics, University of Texas, Austin, TX 78712, USA}
\affiliation[c]{Texas Cosmology Center, University of Texas, Austin, TX 78712, USA}
\emailAdd{elenac@zippy.ph.utexas.edu}
\emailAdd{phn229@physics.utexas.edu}
\emailAdd{jpedraza@physics.utexas.edu}

\abstract{We study the extended thermodynamics, obtained by considering the cosmological constant as a thermodynamic variable, of STU black holes in 4-dimensions in the fixed charge ensemble. The associated phase structure is conjectured to be dual to an RG-flow on the space of field theories. We find that for some charge configurations the phase structure resembles that of a Van der Waals gas: the system exhibits a family of first order phase transitions ending in a second order phase transition at a critical temperature. We calculate the holographic entanglement entropy for several charge configurations and show that for the cases where the gravity background exhibits Van der Waals behavior, the entanglement entropy presents a transition at the same critical
temperature. To further characterize the phase transition we calculate appropriate critical exponents and show that they coincide. Thus, the entanglement entropy successfully captures the information of the {\em extended} phase structure. Finally, we discuss the physical interpretation of the extended space in terms of the boundary QFT and construct various holographic heat engines dual to STU black holes.}
\begin{document}

\maketitle
\flushbottom

\section{Introduction}\label{intro}
The  gauge-gravity correspondence \cite{Maldacena:1997re,Gubser:1998bc,Witten:1998qj} has uncovered deep connections between  black holes in anti-de Sitter space and quantum field theories.  According to the AdS/CFT dictionary, these solutions are the holographic duals to strongly-coupled large-$N$ gauge theories at finite temperature. Since black holes have thermodynamical properties, they seem to undergo phase transitions in the same manner as ordinary thermodynamical systems, and the implications of such phase transition for the boundary theory are worth exploring. The best known example of such a transition is the Hawking-Page transition \cite{HawkingPage}, a first order transition between the Schwarzschild-AdS solution and thermal AdS. The Hawking-Page transition was interpreted in the holographic context by Witten \cite{Witten:1998zw} as the gravity dual to the confinement/deconfinement transition of $\mathcal{N}=4$ SYM.

The study of black hole transitions was then extended to include charged black holes in AdS \cite{Chamblin:1999tk,Cvetic:1999ne}. Compared to the Schwarzschild-AdS case, the phase structure of the Reissner-Nordstr\"om-AdS solution was found to be richer and dependent on the chosen statistical ensemble. In particular, in the canonical ensemble (obtained by keeping fixed the electric charge of the black hole), the behavior of the inverse temperature versus the horizon radius (or, equivalently, the black hole entropy) reveals a phase transition similar to that of a Van der Waals gas. Moreover, there is a critical value of the charge where the system has a second-order critical point.

In a recent series of papers, the analogy with the Van der Waals gas has been made more precise in the framework of extended black hole thermodynamics \cite{Kastor:2009wy,Dolan:2011xt,Dolan:2012jh,Kubiznak:2012wp,Hendi:2012um,Kubiznak:2014zwa,Xu:2014kwa,Grumiller:2014oha,Dolan:2014cja,Dolan:2014jva,Rajagopal:2014ewa,Zhang:2014uoa,Hendi:2014kha,Zhao:2014fea,Zhao:2014owa,Delsate:2014zma,Zhang:2015ova,Hendi:2015kza}. In this context, the cosmological constant is treated as a dynamical variable and is associated with the pressure of the system, $P=-\Lambda/8\pi G_N$, while its conjugate quantity is identified as the thermodynamical volume $V$. The first law is then extended to\footnote{For charged black holes there is an extra term that must also be included into the first law, $dH=T dS+VdP+\Phi dQ$.
However, in the canonical ensemble $dQ=0$ so it reduces to (\ref{firstlaw}).}
\be\label{firstlaw}
dH=T dS+VdP\,,
\ee
where $H$ is the enthalpy of the system (or, equivalently, the ADM mass $M$), $T\equiv\beta^{-1}$ is the black hole temperature and $S$ is the Bekenstein-Hawking entropy,
\be\label{bhent}
S=\frac{A}{4G_N}\,.
\ee
The extended phase space and associated thermodynamics have been studied for the RN-AdS solution, Taub-NUT-AdS, Taub-Bolt-AdS \cite{Johnson:2014xza,Johnson:2014pwa,Lee:2014tma} and their cousins in higher derivative theories such as Gauss-Bonnet and Lovelock gravity \cite{Cai:2013qga,Frassino:2014pha,Dolan:2014vba,Sherkatghanad:2014hda,Hendi:2015oqa,Hennigar:2015esa}. In simple cases such as the RN-AdS solution, the thermodynamical volume is shown to coincide with a naive integration over the black hole interior (in the Schwarzschild slicing),
\be
V=\frac{\Omega_{D-2}r_+^{D-1}}{D-1}\,.
\ee
However, it remains unclear how should it be interpreted, since the volume of a black hole is not a well-defined quantity and depends on the slicing of the spacetime.\footnote{Intriguingly, the thermodynamic volume can be negative in the case of the AdS-Taub-NUT spacetime. In \cite{Johnson:2014xza,Johnson:2014pwa} the volume variable was interpreted in terms of the formation process of the spacetime.} Nevertheless, in most of the cases studied it was found that isotherms on the black hole $PV$ diagrams exhibit the characteristic Van der Waals behavior: the system undergoes a family of first order phase transitions ending in a second order phase transition at a critical temperature.

Unlike the Hawking-Page transition, the Van der Waals transition unfortunately lacks a clear field theory interpretation.
The first steps in this direction were made recently in \cite{Johnson:2014yja}. In this paper, the author proposed that varying the cosmological constant in the bulk corresponds to perturbing the dual CFT, triggering a renormalization group flow. The transition is then interpreted not as a thermodynamical transition but, instead, as transition in the space of field theories. This statement can be intuitively understood since, in holographic theories, the cosmological constant is usually associated to the rank of the gauge group of the dual theory $N$, or its central charge $c$. Thus, changing the value of $\Lambda$ is equivalent to changing the boundary theory. Nonetheless, it is worth emphasizing that the details of the field theory interpretation are still, to a large extent, open questions. In particular, the precise role of the thermodynamical volume $V$ remains to be fully understood.

The Van der Waals transition has also been observed in a different context  \cite{Johnson:2013dka}, although no connection with  extended thermodynamics was pointed out. The authors showed that, like black hole entropy, entanglement entropy also undergoes a Van der Waals-like transition as the temperature varies. However, the cosmological constant was kept fixed, so the emphasis was in finding potential holographic applications to condensed matter systems, \emph{e.g.} \cite{Aprile:2010ge,Aprile:2011uq,Bobev:2011rv}. Here we point out that there is indeed a connection between these two approaches. In particular, we show that the relevant parameter that determines the transition is a dimensionless combination between the temperature and the pressure, $\zeta=\beta^2P$, which can be tuned either by varying the cosmological constant (changing the theory) or the temperature (changing the state).
Since the transition is also visible in the entanglement entropy computation, this result suggests that we can use this field theory observable as an efficient tool to unravel the full space of theories spanned in the $PV$ space.

Before proceeding further, let us motivate the choice of gravity solutions that we will consider throughout this paper.
On the gravity side, the idea of varying the cosmological constant finds its natural setting in theories where the cosmological constant arises as a vacuum expectation value. For example in stringy black hole solutions where the cosmological constant is the minimum of the potential of scalar fields. This was the main motivation for  \cite{Cvetic:2010jb}. In that work, the thermodynamical volume of a class of supergravity solutions called STU black holes was computed and studied in detail. The authors found that for STU black  holes  the thermodynamical volume does not coincide with a naive integration over the black hole interior.\footnote{ Rather, the  thermodynamical volume agrees with the integral of the scalar potential over the black hole interior \cite{Cvetic:2010jb}.} Another advantage of studying the extended phase structure in a specific top-down constructions such as the STU black holes is that the field theory interpretation is easier to understand. In particular, the scalar fields that appear in the bulk system are dual to certain operators in the dual theory that are turned on and off and, therefore, triggering a nontrivial RG-flow in the boundary. Thus, in this work, we will focus on the STU black hole solutions. Other properties of the STU black holes have been explored in \cite{Behrndt:1998jd,Duff:1999gh, Cvetic:1999xp, Gibbons:2005vp}.

In the present work we study  the extended thermodynamics of STU black holes. We first ask ouserlves how generic is the Wan der Waals behavior in these more general backgrounds that involve  four charges. We  find that it is not quite generic; it is not present for all charge configurations. It is then natural to ask whether the holographic entanglement entropy  correctly captures this property. The answer is yes. We show that --even for small entangling regions--  the  holographic entanglement entropy not only distinguishes between STU black holes with and without Van der Waals but also exhibits the same critical temperture and the same  critical exponents as the ones  obtained in the extended thermodynamic framework.

The paper is organized as follows: in Section \ref{STUreview}, we start by briefly reviewing the thermodynamics and phase transitions of STU black holes for different charge configurations. We focus on the case of 4-dimensional black holes, although we suspect that the conclusions must also apply to similar spacetimes in 5 and 7 dimensions. In Section \ref{PVdiagrams}, we study the isotherms on the $PV$ diagram and show that the Van der Waals behavior is observed in two special configurations: the cases of three and four equal charges. In Section \ref{criticalexpsec} we study further the phase transitions and compute the characteristic critical exponents of various thermodynamical variables around the critical point. In Section \ref{Interpretation}, we comment and clarify some previous ideas on the field theory interpretation of the $PV$ phase space and the construction of holographic heat engines. In Section \ref{Holographysec}, we turn to the computation of holographic entanglement entropy. We verify that for charge configurations that exhibit a Van der Waals behavior, the entanglement entropy indeed undergoes a similar transition at the same critical temperature. In Section \ref{cexpee} we compute the critical exponent of the entanglement entropy around the phase transition and show that it is in agreement with the expected result for a Van der Waals gas. Finally, we close in Section \ref{Conclusion}  with conclusions and point some possible directions for future work.

\section{Extended thermodynamics of STU black holes}\label{STUreview}

\subsection{Background and thermodynamical variables}

The STU black hole in 3+1 dimension is a black hole solution of gauged $N=8$, $D=4$ supergravity, first derived in \cite{Duff:1999gh,Cvetic:1999xp}. Gauged $N=8$, $D=4$ supergravity arises by the compactification of $D=11$ supergravity on $S^{7}$, then truncating to the massless sector. The gauge group, which is also the isometry group of the internal space, is $SO(8)$. In order to find black hole solutions, the gauge group is further truncated to its Cartan subgroup $U(1) \times U(1) \times U(1) \times U(1)$. Hence, the black hole solution has four different electric charges.\footnote{Analogous solutions exist in 4+1 and 6+1 dimensions. The 4+1 solution was obtained from compactifying $D=10$ Type IIB supergravity on $S^{5}$. Since the isometry group of $S^{5}$ is $SO(6)$, with Cartan subgroup $U(1) \times U(1) \times U(1)$, the 4+1 black hole has three electric charges. The 6+1 solution was obtained from compactifying $D=11$ supergravity on $S^{4}$. Since the isometry group is $SO(5)$ and its Cartan subgroup is $U(1) \times U(1)$, the 6+1 black hole has two electric charges} The bosonic part of the 4-dimensional effective action reads
:
\begin{equation}
\mathcal{S} = \frac{1}{2\kappa^2}\int d^4x\sqrt{-g}\left(\mathcal{R}-\frac{1}{2}\sum_{i=1}^{3}\left(\partial_{\mu}\phi^{(i)}\right)^{2} - \mathcal{V} - \frac{1}{4} \sum_{I=1}^{4} X_{I}^{-2} F_{\mu\nu}^{(I)2}\right)\,,
\end{equation}
where
\begin{equation}\label{scalar1}
X_{I} = e^{-\frac{1}{2}\vec{a}_{i}\cdot\phi}\,,
\end{equation}
\begin{equation}
\vec{a}_{1} = (1,1,1)\,,
\end{equation}
\begin{equation}
\vec{a}_{2} = (1,-1,-1)\,,
\end{equation}
\begin{equation}
\vec{a}_{3} = (-1,1,-1)\,,
\end{equation}
\begin{equation}
\vec{a}_{4} = (-1,-1,1)\,,
\end{equation}
and the scalar potential is given by
\begin{equation}\label{potdef}
\mathcal{V} = -g^{2} \sum_{i<j} X_{i}X_{j}\,.
\end{equation}
The solution discovered in \cite{Cvetic:1999xp} is given by:
\begin{equation}\label{metric}
ds^{2} = -(H_{1}H_{2}H_{3}H_{4})^{-1/2}fdt^{2} + (H_{1}H_{2}H_{3}H_{4})^{1/2}\left(\frac{dr^{2}}{f} + r^{2}d\Omega^{2}\right)\,,
\end{equation}
\begin{equation}
A^{i} = \frac{\sqrt{q_{i}(q_{i}+2m)}}{r+q_{i}}dt\,,
\end{equation}
\begin{equation}\label{scalar2}
X_{i} = \frac{(H_{1}H_{2}H_{3}H_{4})^{1/4}}{H_{i}}\,,
\end{equation}
where
\begin{equation}\label{eq:STUf}
f = 1 - \frac{2m}{r} + g^2 r^2 H_{1}H_{2}H_{3}H_{4}\,,
\end{equation}
\begin{equation}
H_{i} = 1 + \frac{q_{i}}{r}\,.
\end{equation}
The relevant thermodynamic quantities are given by:
\begin{equation}\label{eq:STUH}
H = m + \frac{1}{4}\sum_{i}q_{i}\,,
\end{equation}
\begin{equation}\label{eq:STUQ}
Q_{i} = \frac{1}{2}\sqrt{q_{i}(q_{i}+2m)}\,,
\end{equation}
\begin{equation}\label{eq:STUS}
S = \pi \prod_{i}\sqrt{r_{+}+q_{i}}\,,
\end{equation}
\begin{equation}\label{eq:STUT}
T = \frac{f'{(r_{+})}}{4\pi} \prod_{i} H_{i}^{-1/2}{(r_{+})}\,,
\end{equation}
\begin{equation}\label{eq:STUPhi}
\Phi^{i} = \frac{\sqrt{q_{i}(q_{i}+2m)}}{2(r_{+}+q_{i})}\,,
\end{equation}
\begin{equation}\label{eq:STUP}
P = \frac{3}{8\pi}g^{2}\,,
\end{equation}
\begin{equation}\label{eq:STUV}
V = \frac{\pi}{3}r_{+}^{3} \prod_{i} H_{i}{(r_{+})} \sum_{j}\frac{1}{H_{j}{(r_{+})}}\,,
\end{equation}
where $H$ is the enthalpy (identified with the ADM mass of the spacetime), $Q_{i}$ are the physical charges, 
$S$ is the entropy, $T$ is the Hawking temperature, $\Phi_{i}$ is the electrostatic potential at infinity, $P$ is the pressure and $V$ is the thermodynamic volume. Notice that we are working in Planck units ($c=\hbar=k_{B}=G=1$), so all quantities are dimensionless. The outer horizon is located at $r=r_{+}$, the largest root of $f(r_+)=0$.

For the STU black hole, the first law of thermodynamics reads:
\begin{equation}
dH = TdS + \sum_{i=1}^{4} \Phi_{i}dQ_{i} + VdP\,.
\end{equation}
In the remainder of the paper we will work in the canonical ensemble, \emph{i.e.} we keep all the physical charges $Q_{i}$ fixed, so the first law simplifies to:
\begin{equation}\label{eq:firstlaw_fixedq}
dH = TdS + VdP\,.
\end{equation}
We can easily check that the Hawking temperature (\ref{eq:STUT}) and the thermodynamic volume (\ref{eq:STUV}) can be recovered from the enthalpy by the usual thermodynamic relations:
\begin{equation}
T = \left(\frac{\partial H}{\partial S}\right)_{P}\,,
\end{equation}
\begin{equation}
V = \left(\frac{\partial H}{\partial P}\right)_{S}\,.
\end{equation}

\subsection{Extended phase structure}\label{PVdiagrams}
We will now proceed to study some aspects of the extended phase structure of STU black holes. In order to fix notation and to illustrate the procedure we will first review the special case of AdS-RN which have already been studied in various papers, \emph{e.g.} \cite{Johnson:2014yja}.

\subsubsection{The AdS-RN black hole}
An AdS-RN black hole is a special case of an STU black hole when all the charges are equal. Setting $q_1=q_2=q_3=q_4 \equiv q$ in (\ref{metric})
we obtain,
\begin{equation}\label{eq:metricAdSRN}
ds^{2} = -\left(1+\frac{q}{r}\right)^{-2}fdt^{2} + \left(1+\frac{q}{r}\right)^{2}\left(\frac{dr^{2}}{f} + r^{2}d\Omega^{2}\right)\,,
\end{equation}
where
\begin{equation}\label{eq:fAdSRN}
f=1-\frac{2m}{r} + g^2\left(1+ \frac{q}{r}\right)^4 r^2\,.
\end{equation}
The thermodynamic quantities entering (\ref{eq:firstlaw_fixedq}) are now,
\begin{align}
H&=m+q\,, \label{eq:HAdSRN}\\
T&=\frac{m r_+-g^2(q-r_{+})(q+r_{+})^3}{2 \pi r_{+}(q+r_{+})^2}\,,\label{eq:TAdSRN}\\
S&=\pi(q+ r_{+})^2\,,\label{eq:SAdSRN}\\
V&=\frac{4}{3}\pi (q+r_{+})^3\,, \label{eq:VAdSRN}\\
P&=\frac{3}{8\pi} g^2\,.\label{eq:PAdSRN}
\end{align}

A peculiarity of the AdS-RN solution is that the pressure and volume determine each other. Indeed, from equations (\ref{eq:SAdSRN}) and (\ref{eq:VAdSRN}) it follows that:
\begin{equation}\label{eq:SVAdSRN}
V = \frac{4}{3\sqrt{\pi}}S^{3/2}\,.
\end{equation}
One might be worried about this, since it means that the volume and entropy cannot be used as independent thermodynamical variables. Fortunately this entropy-volume degeneracy is specific to AdS-RN, volume and entropy are independent variables in the other STU special cases (as well as for rotating black holes, for instance). Also recall that the physical charge is not $q$ but
\begin{equation}\label{eq:QAdSRN}
Q=\frac{1}{2}\sqrt{q(q+2m)}\,.
\end{equation}

At this point would be convenient to introduce the radial Schwarzschild coordinate $R=r+q$ which would allow us to write the metric in a more familiar form\footnote{In these coordinates the AdS-RN metric takes the form
\begin{equation*}
ds^{2} = -F{(R)}dt^{2} + \frac{dR^{2}}{F{(R)}} + R^{2}d\Omega^{2}\,,
\end{equation*}
with
\begin{equation*}
F{(R)} = 1 - \frac{2M}{R} + \frac{4Q^{2}}{R^{2}}+g^{2}R^{2}\,.
\end{equation*}}
and obtain the equation of state in a straight forward manner. However, in the general case such strategy is not available and one has to proceed numerically in order to obtain the equation of state. Therefore, we will refrain here from switching to the Schwarzschild coordinate here and work directly with equations (\ref{eq:metricAdSRN})-(\ref{eq:PAdSRN}).
From (\ref{eq:VAdSRN}) and (\ref{eq:fAdSRN}) we  obtain  horizon and mass parameter as a function of $q$, $V$ and $P$:
\begin{align}
r_+&= -q + \frac{1}{2}\left(\frac{6}{\pi}\right)^{1/3} V^{1/3}\,,\label{eq:rhVq}\\
m&=-\frac{q}{2} + \frac{1}{4}\left(\frac{6}{\pi}\right)^{1/3} V^{1/3}\left(1+\frac{4 P\ V }{-2 q + \left(\frac{6 V}{\pi}\right)^{1/3}}\right)\,.\label{eq:mPVq}
\end{align}
Using  \eqref{eq:mPVq}, \eqref{eq:PAdSRN}  and \eqref{eq:QAdSRN}  we can then solve for the charge parameter $q$ as a function of the physical charge $Q$ , the volume $V$ and the pressure $P$.

\begin{figure}
\center{
\includegraphics[width=8.5cm]{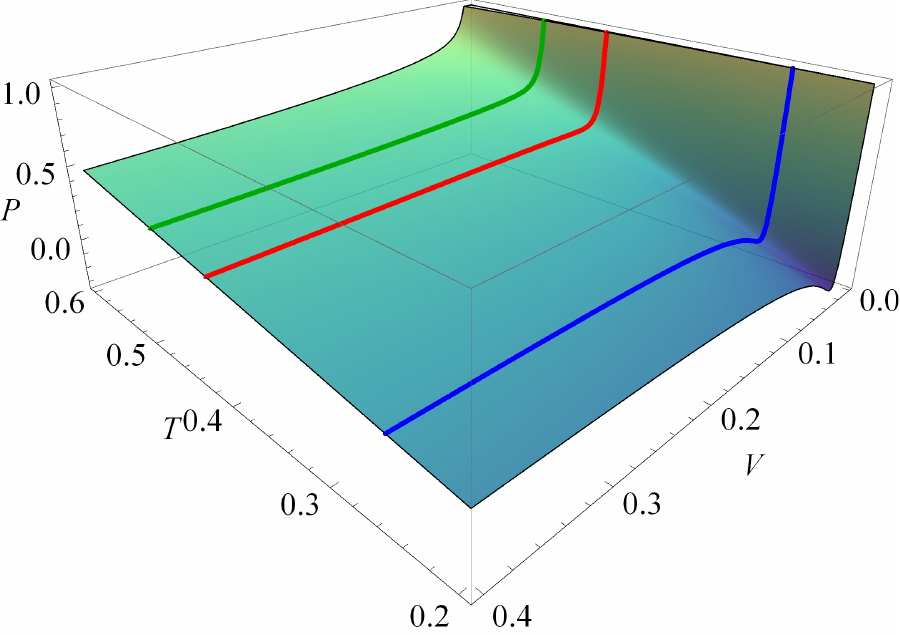}}
\vspace{-0.2cm}
\caption{$PVT$ diagram for AdS-RN with charge $Q=0.05$. The solid color curves are isotherms. The blue one clearly exhibits Van der Waals behavior but the green one does not. The critical temperature at which the transition occurs is depicted in red, and can be determined from the conditions (\ref{infpoint}).\label{fig:3dPVT}}
\end{figure}

With all these ingredients at hand, the temperature \eqref{eq:TAdSRN} defines the equation of state, $T=\mathcal{F}(Q,V,P)$,
which we can rewrite as
\begin{equation}\label{RNAdSeos}
P = \frac{1}{8\pi}\left(\frac{4\pi}{3}\right)^{4/3}\left(\frac{3T}{V^{1/3}}-\left(\frac{3}{4\pi}\right)^{2/3}\frac{1}{V^{2/3}} + \frac{4Q^{2}}{V^{4/3}} \right)\,.
\end{equation}
As shown in Figure \ref{fig:3dPVT}, this equation of state exhibits a Van der Waals-like phase transition at sufficiently low temperature. 
From this plot, we can see that there is a critical temperature, $T_c$, at which there is an inflection point with
\begin{equation}\label{infpoint}
\left(\frac{\partial P}{\partial V}\right)_{T_{c}} = \left(\frac{\partial^{2} P}{\partial V^{2}}\right)_{T_{c}} = 0\,.
\end{equation}
This corresponds to a second order phase transition, above which there is no longer a Van der Waals transition. The critical pressure, volume and temperature are given by
\begin{equation}\label{critvalues}
P_{c} = \frac{1}{384\pi Q^{2}}\,,\qquad V_{c} = 64\sqrt{6}\pi Q^{3}\qquad\text{and}\qquad\beta_{c} = 6\sqrt{6}\pi Q\,,
\end{equation}
respectively.

Before proceeding further let us note that it may appear surprising that the dual field theory, which lives on a finite volume, could undergo a phase transition at all. Phase transitions require non-analyticities in the partition function. Such non-analyticities only arise in system with infinitely many degrees of freedom, which, in more familiar thermodynamical systems, implies infinite volume. In our case, there are indeed an infinite number of particles since we are working in the large-$N$ limit, hence the possibility of phase transitions.

Notice that  Van der Waals behavior of AdS-RN was first discovered  \cite{Chamblin:1999tk} without considering the cosmological constant as a thermodynamical variable. Indeed, if we do not consider the extended thermodynamics and  work at a fixed pressure we arrive to the following equation of state:
\begin{equation}
T = \frac{1}{4\pi } \left(3 g^2 \sqrt{\frac{S}{\pi}} + \sqrt{\frac{\pi}{S}} - 4Q^{2}\left(\frac{\pi}{S}\right)^{3/2}\right)\,.
\end{equation}
It is possible to show that there exists a critical value of the charge $Q_{c}$ at which this curve has an inflection point
\begin{equation}
\left(\frac{\partial \beta}{\partial S}\right)_{Q_{c}} = \left(\frac{\partial^{2} \beta}{\partial S^{2}}\right)_{Q_{c}} = 0\,,
\end{equation}
for which one finds
\begin{equation}\label{ScRNAdS}
S_{c} = \frac{\pi}{6 g^2}\,,\qquad Q_{c} = \frac{1}{12 g}\qquad\text{and}\qquad\beta_{c} = \sqrt{\frac{3}{2}}\pi \frac{1}{g}\,.
\end{equation}
For $Q>Q_{c}$, the curve is thermodynamically stable, while for $Q<Q_{c}$, the plot develops an unstable branch in addition to two stable branches. Also, note that if we plug $g(Q_c)$ in $\beta_{c}$ we recover the same critical temperature as in (\ref{critvalues}). This indicates that the relevant (dimensionless) parameter that determines the transition is $\xi=\beta g=(TL)^{-1}$, which can be tuned either by varying the AdS radius $L$ (\emph{i.e.} the cosmological constant) or the temperature $T$.
\begin{figure}
$$
\begin{array}{cc}
 \includegraphics[width=7.5cm]{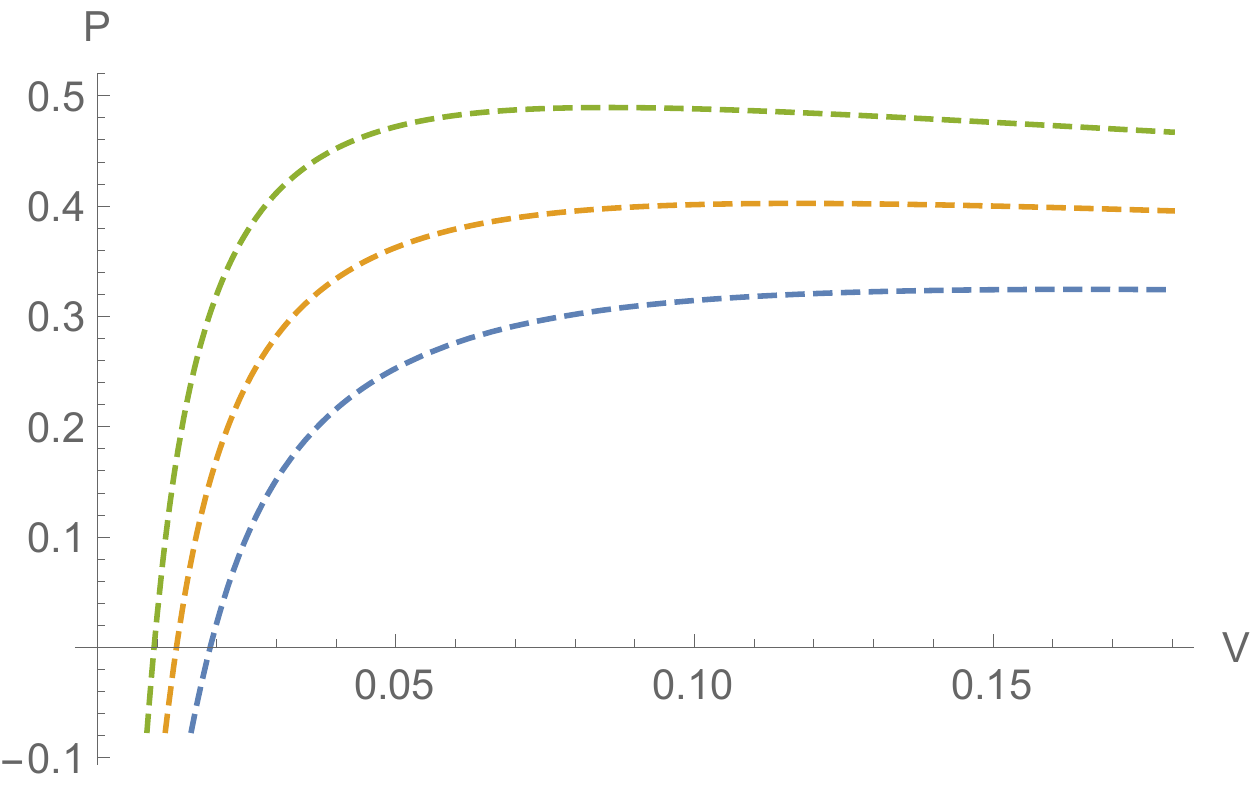} & \includegraphics[width=7.5cm]{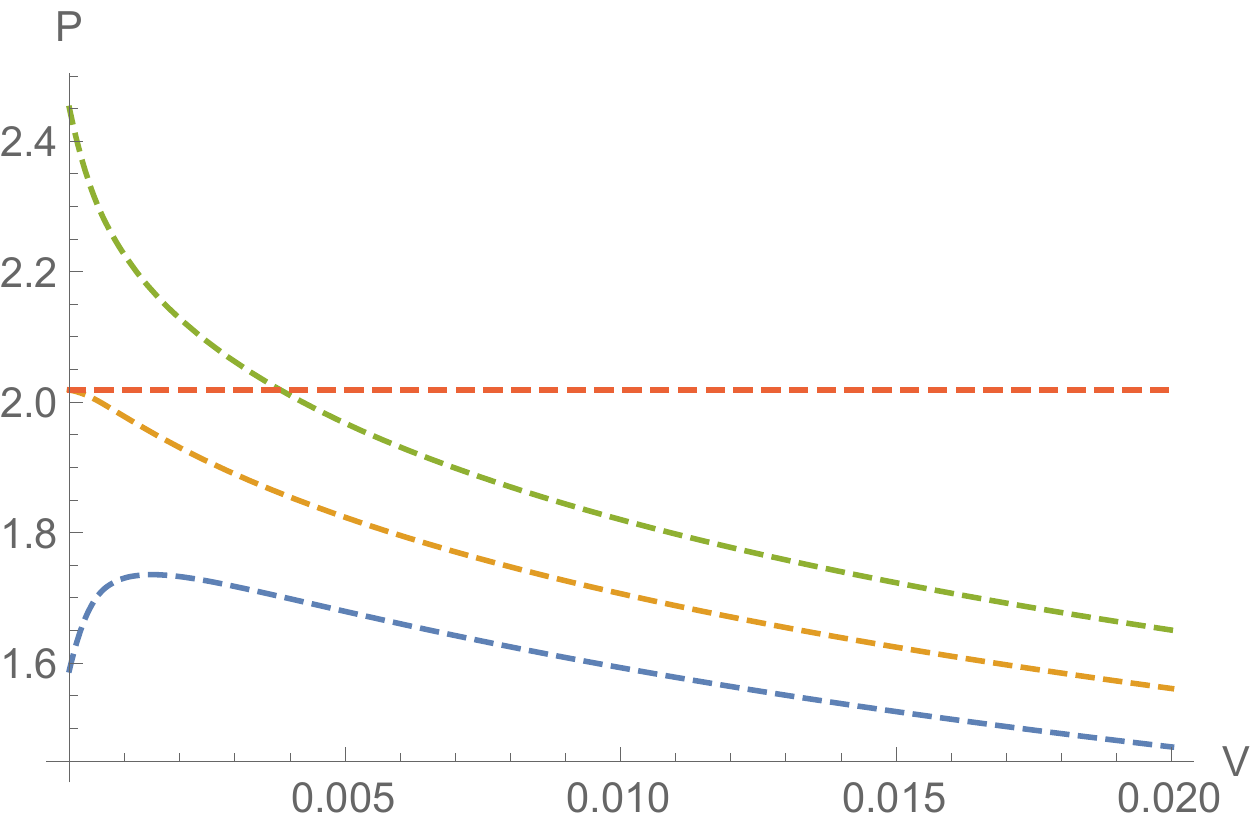}\\
 \quad\quad(a) & \quad\quad(b)\\
  \includegraphics[width=7.5cm]{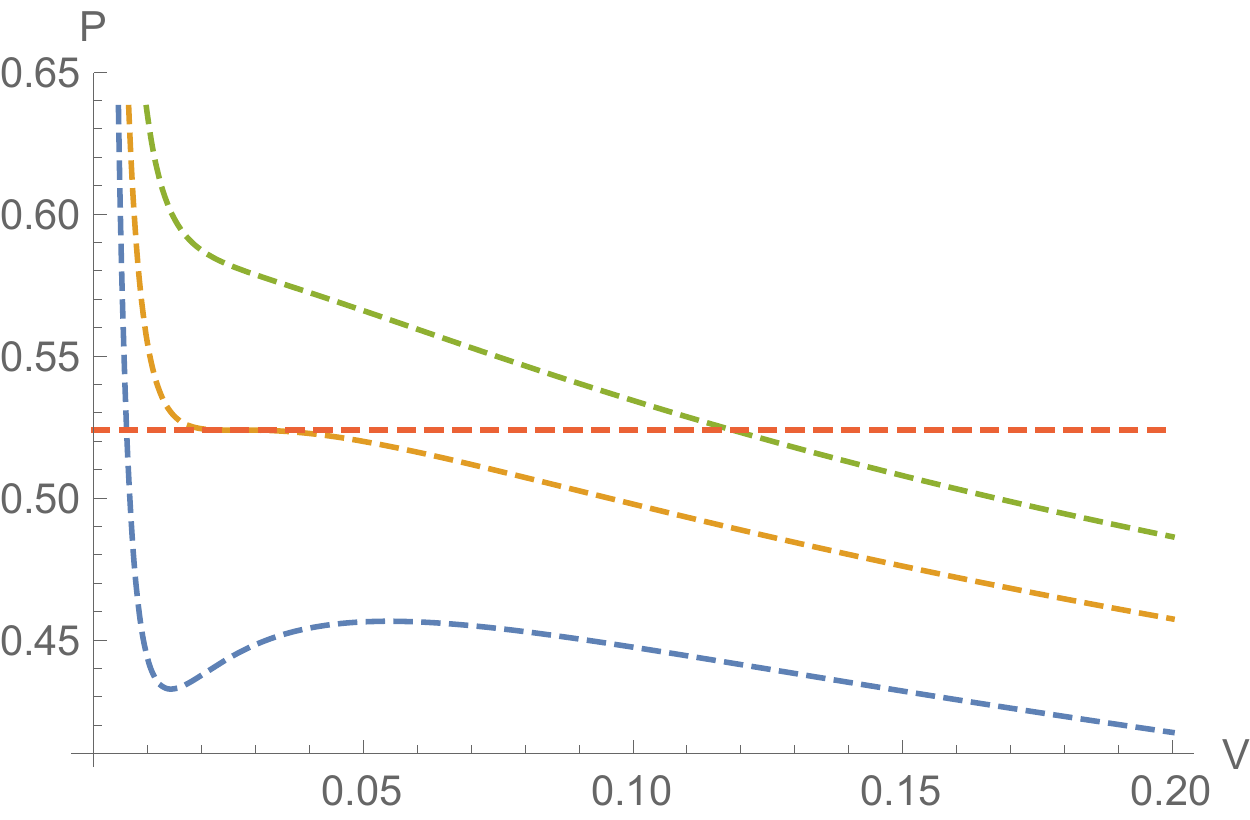} & \includegraphics[width=7.5cm]{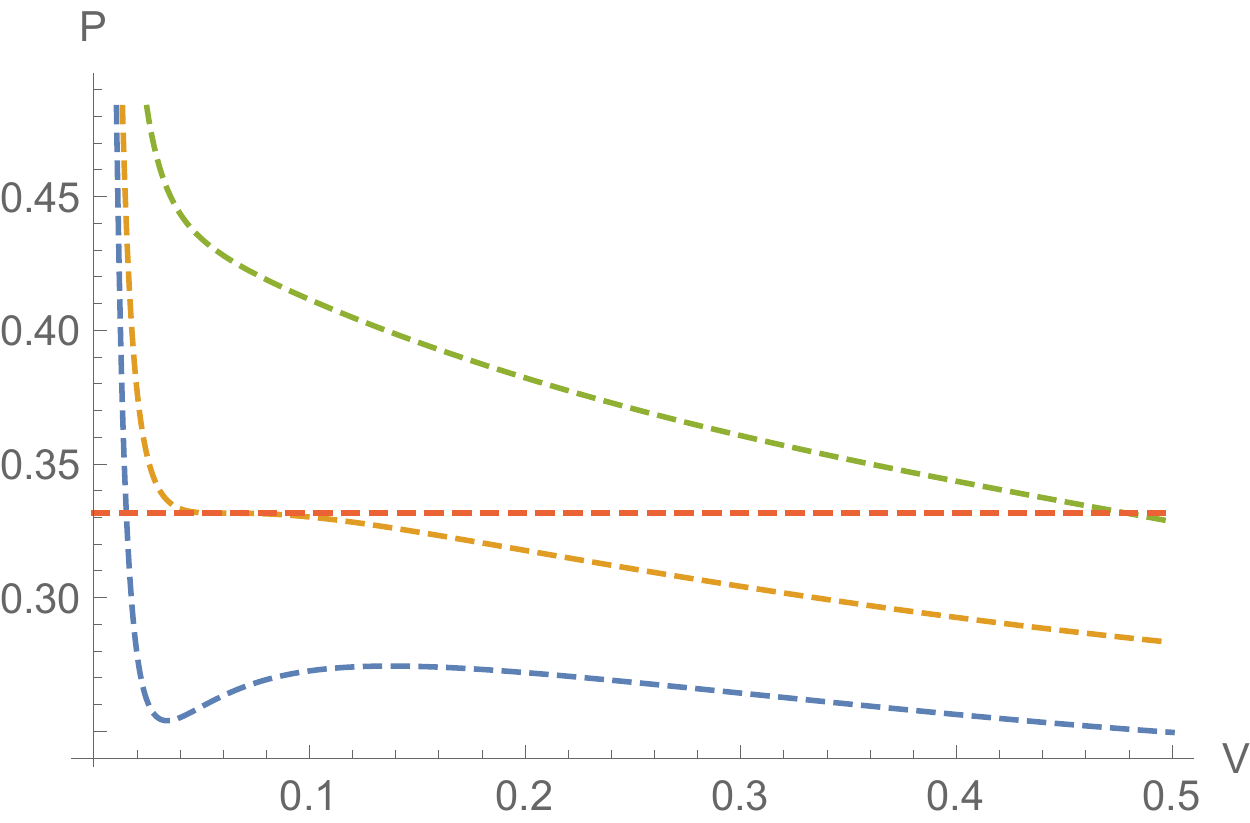}\\
 \quad\quad(c) & \quad\quad(d)\\
\end{array}
$$
\vspace{-0.5cm}
\caption{Isotherms in the $PV$ diagrams of the STU black hole in 3+1 dimensions. Top left: The single charge case $Q_{1}=0.05$ and $Q_{2}=Q_{3}=Q_{4}=0$. The temperatures chosen are $T = 0.55$ (green), $0.5$ (orange) and $0.45$ (blue). Top right: The two-charge case $Q_{1}=Q_{2}=0.05$, $Q_{3}=Q_{4}=0$. The temperatures chosen are $T= 0.96$ (green), $0.93$ (orange) and $0.90$ (blue). Bottom left: The three-charge case $Q_{1}=Q_{2}=Q_{3}=0.05$, $Q_{4}=0$. The temperatures chosen are $T=0.56$ (green), $0.5391$ (orange) and $T=0.51$ (blue). Bottom right: The AdS-RN case with $Q_{1}=Q_{2}=Q_{3}=Q_{4}=0.05$. The temperatures chosen are $T=0.48$ (green), $0.43316$ (orange) and $0.4$ (blue). For the two-charge, three-charge and AdS-RN cases, the orange curve is at critical temperature.}
\label{PVdiagram4dfig}
\end{figure}

\subsubsection{Other charge configurations}

Our goal here  is to repeat the analysis outlined in the previous subsection for  different charge configurations. The logic and the steps are the same but once we move away from the AdS-RN case it is not possible to find an analytic expression for the equation of state. Thus, we have to proceed numerically.

We will focus on cases where one, two or three charges are turned on but, for the sake of simplicity, we will consider all of them to be equal, \emph{i.e.} we will set $q_i=q$ $\,\,\forall\, i$. The thermodynamic quantitites are given in \eqref{eq:STUQ}-\eqref{eq:STUV}. In particular, the temperature for a general STU black hole is \eqref{eq:STUT},
\begin{equation}\label{eq:STUtemperature}
T=\frac{2  P}{3 \sqrt{\prod_{i=1}^4 (1+\frac{q_i}{r_+})}} \left[- \sum_{i=1}^4 q_i\prod_{j\ne i}\left(1+ \frac{q_j}{r_+}\right) +\frac{2 m}{ r_+^2}\frac{3}{8\pi P} + 2  r_+\prod_{i=1}^4 \left(1+\frac{q_i}{r_+}\right) \right].
\end{equation}

Using \eqref{eq:STUQ}, \eqref{eq:STUV} and \eqref{eq:STUf} one can solve for $r_+,\ m$ and $q_i$ in terms of $Q_i$ and $V$. Thus, \eqref{eq:STUtemperature}
implicitly  defines the equation of state relating  $T,\ P,\ V$ and the physical charge $Q$.
In all cases studied it is  possible to obtain three dimensional $PVT$  plots similar to Figure \ref{fig:3dPVT}. However, for maximum clarity we choose to plot constant temperature curves in a $PV$ diagram.

The numerical results are shown in Figure \ref{PVdiagram4dfig}, for different number of charges. As mentioned before, we only consider the cases where the nonzero charges are all equal to each other. However, our numerical explorations indicate that the behavior is qualitatively similar if the nonzero charges are not equal to each other. That is to say, the AdS-RN case will be representative of all cases where all four charges are nonzero, etc. As can be seen from the plots, the Van der Waals behavior is present in the case of three and four active charges, \emph{i.e.} below a certain critical temperature $T_{c}$ the isotherm has two stable branches (where the volume decreases with increasing pressure) and one unstable branch (where the volume increases with increasing pressure). Moreover, for the three-charge case we can use the 3-dimensional plot $P{(V,T)}$ to find $T_{c}$, $P_{c}$ and $V_{c}$ numerically. For $Q = 0.05$, the results are:
\begin{equation}
(T_{c},V_{c},P_{c}) = (0.539,0.0264,0.524)\,.
\end{equation}
But from dimensional analysis, we know that:
\begin{equation}
T_{c} \propto \frac{1}{Q}\,,\qquad V_{c} \propto Q^{3}\,,\qquad P_{c} \propto \frac{1}{Q^{2}}\,,
\end{equation}
where the proportionality factors are dimensionless numbers. Therefore, we can easily deduce the critical temperature, volume and pressure for arbitrary $Q$:
\begin{equation}
T_{c} = \frac{0.0270}{Q}\,,\qquad V_{c} = 211.254 Q^{3}\qquad\text{and}\qquad P_{c} = \frac{0.00131}{Q^{2}}\,,
\end{equation}
respectively. These are equivalent expressions to (\ref{critvalues}), for the case of the AdS-RN black hole. Its worth emphasizing that if we change the value of some of the charges, the constants of proportionality will change, but the phase structure will be similar.

The Van der Waals behavior is absent in the two other cases (one or two active charges). In the case of two active charges, at sufficiently low temperatures, the isotherms has an unstable branch at small volumes and a stable branch at large volumes. As the temperature increases, the system undergoes a phase transition at some critical temperature where the unstable branch is squeezed out. The critical temperature at which this happens is $T_{c} = 0.0465 Q^{-1}$. Finally, in the case of only one active charge, there is an unstable branch for small volume and a stable branch at large volumes, and the unstable branch is present at all temperatures. This is the same behavior as for the zero-charge case (\emph{i.e.} Schwarzschild-AdS), so there must also be a transition at small volumes reminiscent of a Hawking-Page transition, presumably to a fixed charge ensemble in pure AdS.

Next, we turn our attention to plots of the temperature versus the black hole entropy.
\begin{figure}[t!]
$$
\begin{array}{cc}
 \includegraphics[width=7.5cm]{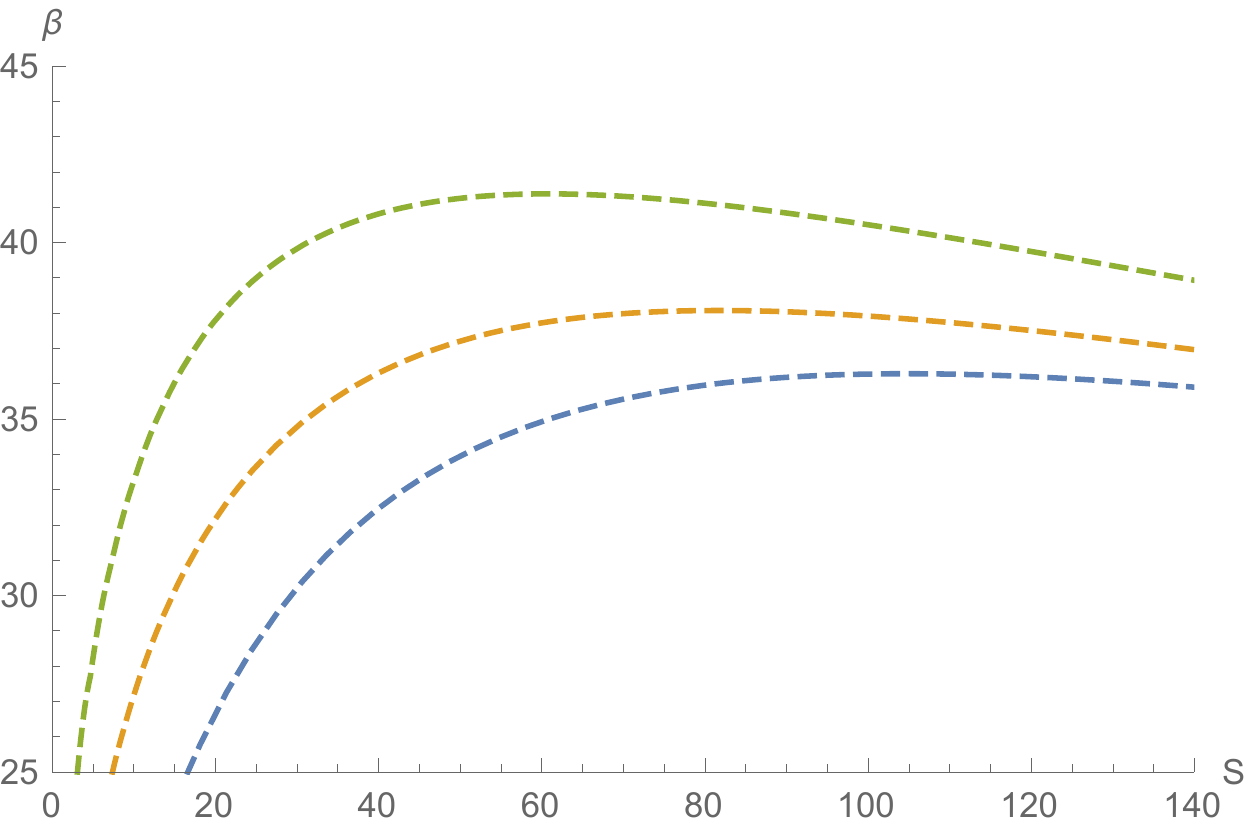} & \includegraphics[width=7.5cm]{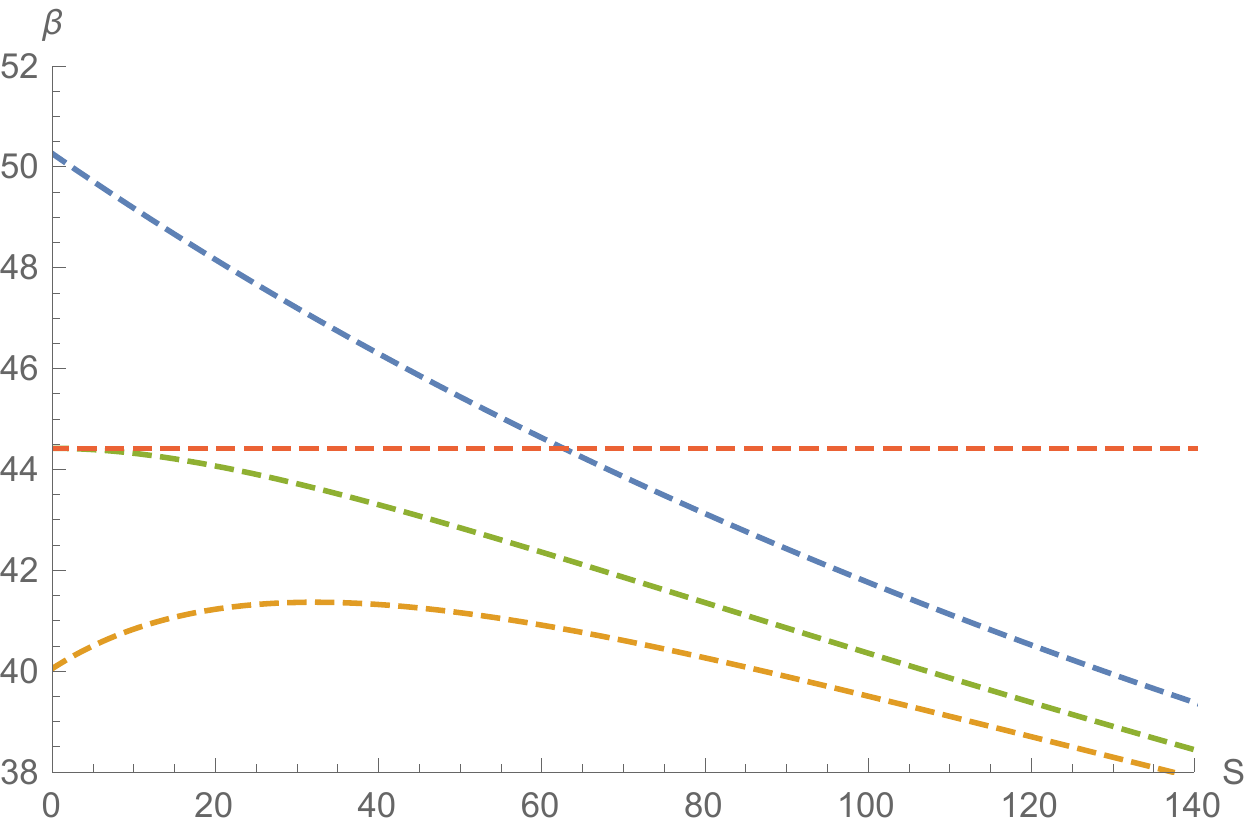}\\
 \quad\quad(a) & \quad\quad(b)\\
  \includegraphics[width=7.5cm]{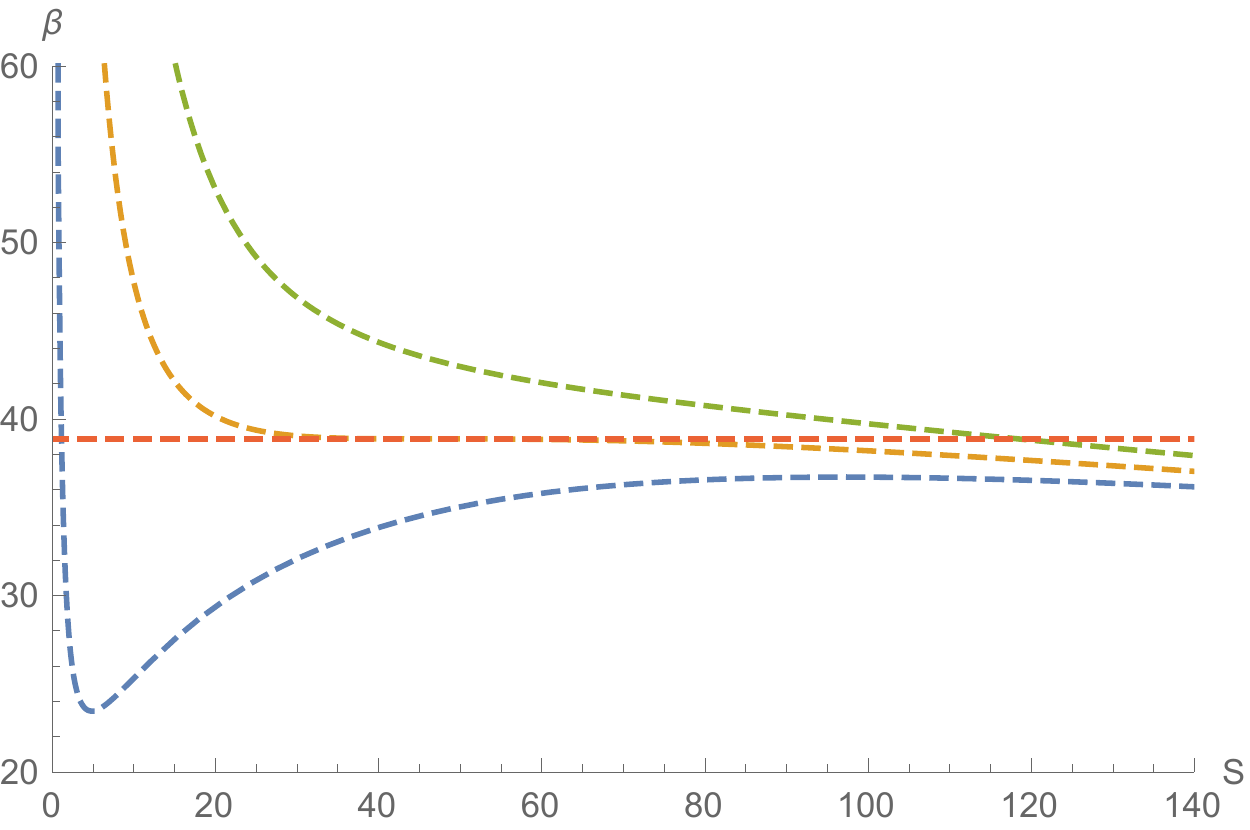} & \includegraphics[width=7.5cm]{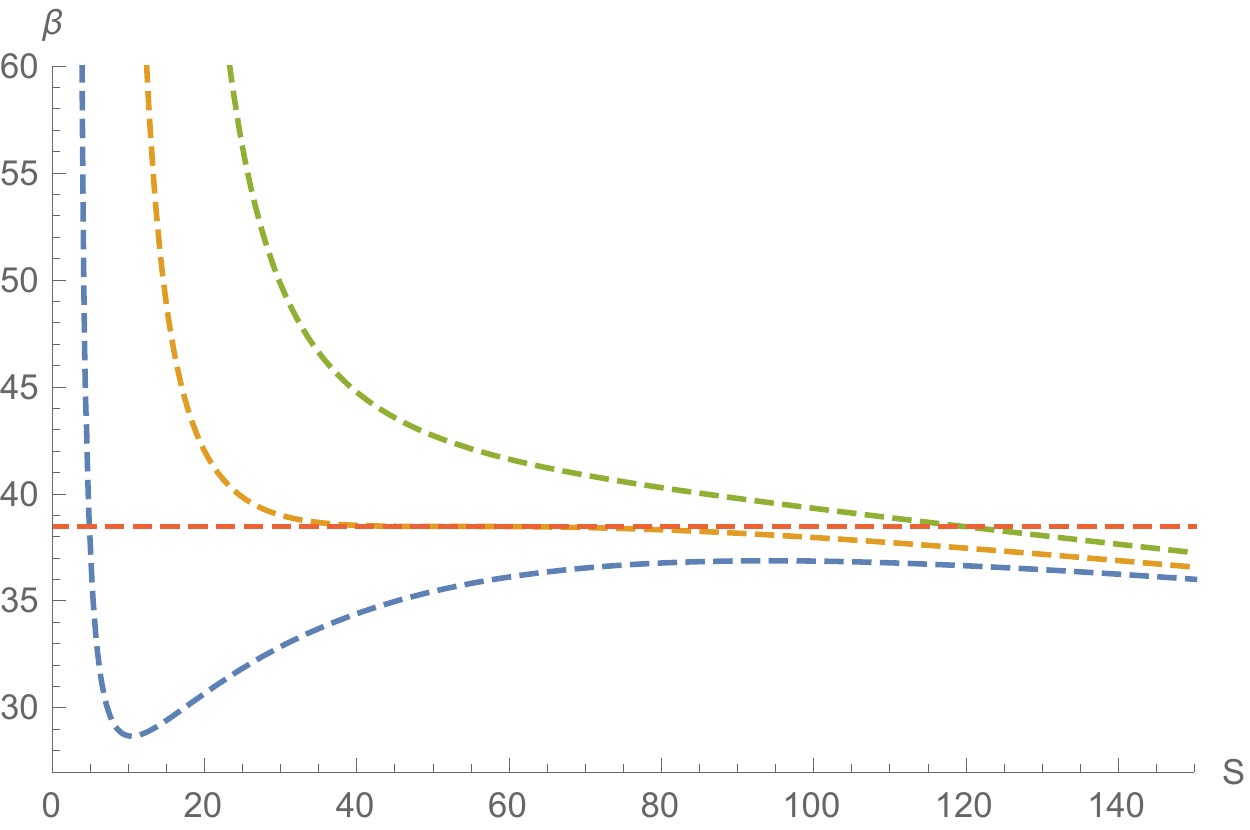}\\
 \quad\quad(c) & \quad\quad(d)\\
\end{array}
$$
\vspace{-0.5cm}
\caption{Plots of the inverse temperature as function of the black hole entropy $S$ for the one-charge case (top left), two-charge case (top right), three-charge case (bottom left) and AdS-RN case (bottom right). For each panel, each of the 3 curves correspond to a fixed numerical value of the common charge. In the upper two plots, the middle curve corresponds to the critical charge and the red horizontal line indicates the critical temperature.}
\label{betavsSfig}
\end{figure}
In Figure \ref{betavsSfig}, we present these plots for the four different cases of the STU black hole that we are considering. Each curve corresponds to a different value of the charge, while keeping the AdS radius fixed $L=10$. As previously, we have generated 3-dimensional plots of $(S,\beta,Q)$ but we have chosen to present 2-dimensional slices for clarity.
Remarkably, the plots are qualitatively the same as Figure \ref{PVdiagram4dfig}: for the three-charge and four-charge cases, the curves have an unstable middle branch at sufficiently low temperature.\footnote{This branch is indeed unstable, since its specific heat $C_P = T\left(\frac{\partial S}{\partial T}\right)_P$ is negative.} We indicate the critical temperature in each case by a red horizontal line. For the AdS-RN case, the critical charge and critical temperature follow from (\ref{ScRNAdS}):  $Q_{c} = 5/6$ and $\beta_{c} = 10\sqrt{3/2}\pi$. For the three-equal-charge case, we made use of the 3-dimensional plot to solve for the critical temperature, entropy and charge:
\begin{equation}
(Q_{c},\beta_{c},S_{c}) = (1.048, 38.871, 46.959)\,.
\end{equation}
These values are valid for $g=1/L=0.1$. From dimensional analysis, we know that:
\begin{equation}
Q_{c} \propto \frac{1}{g}\,,\qquad\beta_{c} \propto \frac{1}{g}\,,\qquad S_{c} \propto \frac{1}{g^2}\,.
\end{equation}
Therefore, we can deduce the formula for the critical charge, temperature and entropy at arbitrary pressure:
\begin{equation}
Q_{c} = \frac{0.1048}{g}\,,\qquad \beta_{c} = \frac{3.8871}{g}\qquad\text{and}\qquad S_{c} = \frac{0.46959}{g^2}\,,
\end{equation}
respectively. On the other hand, the two-charge and single-charge cases do not exhibit the Van der Waals transition. The two-charge case also undergoes a phase transition where the unstable branch is shrunk to zero. This happens at the critical charge $Q_{c} \approx 0.20704/g$ and critical temperature $\beta_{c} \approx 4.44187/g$. Finally, for the one-charge case, the unstable branch persists at all temperatures.

It is worth pointing out that for charge configurations that exhibit a transition, the parameter $\xi=\beta g=(TL)^{-1}$ is fixed in both scenarios, \emph{i.e.} in the $PV$ diagram at constant $T$ or in the $TS$ diagram at constant $P$. This result shows that we can study the phase transition either by tuning the pressure or the temperature, with completely different interpretations in terms of the dual field theory. We will comment more on this point in Section \ref{Interpretation}.

\subsection{Characterization of the phase transitions\label{criticalexpsec}}
An important characterization of continuous phase transitions is given by the so-called critical exponents, which determine the leading power dependence of various quantities close to the phase transition. Remarkably, thermodynamical systems are often found with identical critical exponents even though their microscopic details may differ. This suggests that the critical exponents may be used to classify phase transitions into universality classes. In particular, from the equation describing the critical isotherm $P{(T_{c},V)}$, we can extract the critical exponent $\delta$ defined to be:
\begin{equation}\label{deltadef}
|P-P_{c}| \propto |V-V_{c}|^{\delta}\,.
\end{equation}
The factor of proportionality in the equation above depends on the microscopic details of the system, and therefore is unimportant. For the AdS-RN case equation (\ref{deltadef}) takes the form
\begin{equation}
|P(T_{c},V) - P_{c}| = \frac{1}{2916\sqrt{6}\pi^{4}(4Q)^{11}}|V-V_{c}|^{3}\,,
\end{equation}
and $\delta$ is found to be \cite{Kubiznak:2012wp}:
\begin{equation}
\delta = 3\,.
\end{equation}
Next, we consider the specific heat at constant pressure $C_{P}$, defined by:
\begin{equation}
C_{P} = T\left(\frac{\partial S}{\partial T}\right)_{P}\,.
\end{equation}
We will denote $\alpha$ as the critical exponent associated with $C_{P}$, which at leading order behaves as
\begin{equation}\label{alphadef}
C_{P} \propto (T - T_{c})^{-\alpha}\,.
\end{equation}
For AdS-RN, we have the expansion:
\begin{equation}
|\beta(S,Q_{c}) - \beta_{c}| = 192\sqrt{\pi}P^{5/2}(S-S_{c})^{3}\,,
\end{equation}
and after some algebra, we find that the numerical value of $\alpha$ is given by \cite{Johnson:2013dka}
\begin{equation}
\alpha = \frac{2}{3}\,.
\end{equation}
Note that, since we are working in the canonical ensemble, the specific heat we refer to as $C_P$ is also at constant charge $Q$. More generally, for generically chosen paths $Q(T)$ approaching the critical point, the critical exponent associated to the specific heat $C_P$ its found to be $\gamma=1$ \cite{Kubiznak:2012wp}. It is only for the special case we consider here, $Q=Q_c$, that the exponent is given by 2/3. Remarkably enough, these two critical exponents happen to agree with the expected values for a Van der Waals gas.\footnote{
Our $\alpha$ is not to be confused with the critical exponent usually denoted by $\alpha$ in statistical mechanics. This latter is associated with $C_{V}$.} In addition, several other critical exponents for AdS-RN as well as for most AdS black holes also agree with expectations from mean-field theory \cite{Kubiznak:2012wp}. The idea here is to perform a similar study for the STU black holes.

\begin{figure*}[t!]
$$
\begin{array}{cc}
 \includegraphics[width=7.5cm]{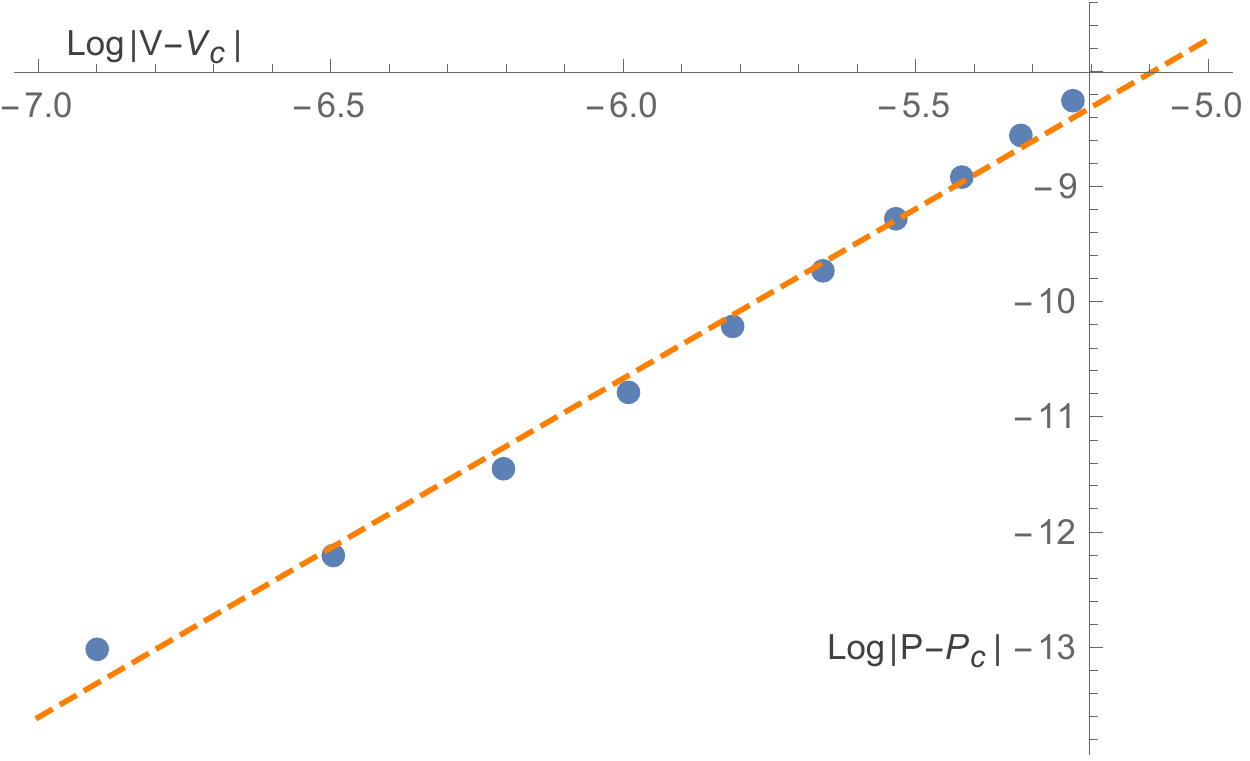} & \includegraphics[width=7.5cm]{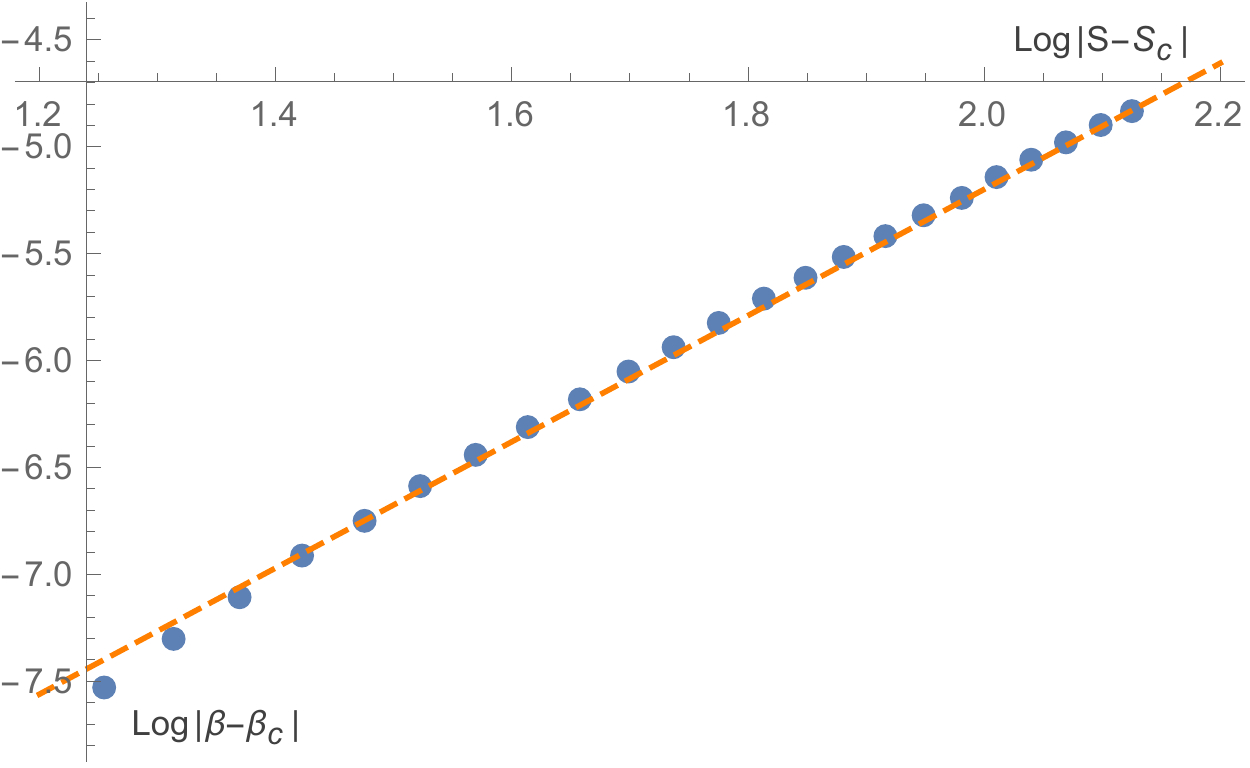}\\
 \quad\quad(a) & \quad\quad(b)\\
\end{array}
$$
\vspace{-0.5cm}
\caption{Left panel: $\log{|P-P_{c}|}$ vs. $\log{|V-V_{c}|}$ for the three-equal-charge black hole with $Q=0.05$. The linear fit is based on 10 points closest to the inflection point, in the volume range $V \in [0.0210,0.0254]$. The critical volume is at $V_{c} = 0.0264$. Right panel: $\log{|\beta-\beta_{c}|}$ vs. $\log{|S-S_{c}|}$ for the three-equal-charge black hole with $Q=0.05$. The linear fit is based on 23 points closest to the inflection point, in the entropy range $S \in [50.4622, 55.3306]$.}
\label{DeltaGammaplot}
\end{figure*}

Recall that for other charge configurations we do not have analytical expressions for the equation of state so we have to proceed numerically. We will focus on the three-equal-charge black hole, which is the only other case in consideration that exhibits a Van der Waals-like transition ---see Figure \ref{PVdiagram4dfig}.
In order to extract the value of $\delta$ we first take the logarithm of both sides of (\ref{deltadef}),
\begin{equation}
\log{|P-P_{c}|} = \delta\log{|V-V_{c}|} + C\,,
\end{equation}
where $C$ is an additive constant. Therefore, we can plot $\log{|P-P_{c}|}$ as a function of $\log{|V-V_{c}|}$ in the vicinity of $V_{c}$ and as a result we expect a straight line with slope $\delta$. We present this plot in Figure \ref{DeltaGammaplot}(a), with the common value of the charges set to $Q = 0.05$.
A linear fit of the data points yields:
\begin{equation}\label{deltanumerical}
\log{|P-P_{c}|} = 2.9476\log{|V-V_{c}|} + 7.0200\,,
\end{equation}
with a slope that suggests the exact value:
\begin{equation}
\delta = 3\,.
\end{equation}
Notice that this is the same value that we found for AdS-RN. For comparison, the analog of equation (\ref{deltanumerical}) for AdS-RN with four all charges set to $Q=0.05$ reads:
\begin{equation}
\log{|P-P_{c}|} = 3\log{|V-V_{c}|} + 4.25105\,.
\end{equation}
The $y$-intercept of the equation above differs from that of (\ref{deltanumerical}), even though they are of the same order of magnitude. This difference reflects the different microscopic details between AdS-RN and the three-equal-charge STU black hole.

Next, we can repeat the same procedure for the $TS$ diagrams, and consider the power dependence of $S$ around the critical point (on the critical charge curve). To do so, we plot $\log{|\beta-\beta_{c}|}$ vs. $\log{|S-S_{c}|}$ in Figure \ref{DeltaGammaplot}(b), with $g=0.1$. A linear fit of this curve yields:
\begin{equation}\label{alphanumerical}
\log{|\beta-\beta_{c}|} = 3.0486\log{|S-S_{c}|} - 11.2609\,,
\end{equation}
suggesting that
\begin{equation}
\beta - \beta_{c} \propto |S-S_{c}|^{3}\,.
\end{equation}
Then, from (\ref{alphadef}) it follows that
\begin{equation}
\alpha = \frac{2}{3}\,.
\end{equation}
Again we find the same value as the AdS-RN case. For comparison, the analog expression to (\ref{alphanumerical}) for AdS-RN, with $g=0.1$, is:
\begin{equation}
\log{|\beta-\beta_{c}|} = 3\log{|S-S_{c}|} - 10.997\,.
\end{equation}

\subsection{Field theory interpretation\label{Interpretation}}

The analysis of \cite{Johnson:2014yja} lead to some puzzles regarding the implications of the extended thermodynamics in the context of the AdS/CFT correspondence. Here we will review the main issues and discuss the resolutions that have been proposed in terms of a specific top-to-bottom construction, where the physical interpretation is neater.

According to the holographic dictionary, $(3+1)$-dimensional asymptotically AdS black holes are dual to CFTs at finite temperature in $2+1$ dimensions. However, the pressure and volume appearing in the extended thermodynamics are \emph{not} identified with the pressure and volume of the dual field theory, instead, they are interpreted as (non-thermodynamic) dynamical variables in the space of field theories \cite{Johnson:2014yja}.

Let us first recall the main entries in the path integral formulation of AdS/CFT. For a fixed cosmological constant $\Lambda$, the Euclidean path integral in the gravity side,
\be\label{IEgravity}
I^E/\beta=-\log \mathcal{Z} /\beta\,,
\ee
is equivalent to the Helmholtz free energy $F=U-TS$ of the dual CFT. However, in top-to-bottom constructions (as the one we consider here), the value of $\Lambda$ is set by the potential $\mathcal{V}(\phi^i)$ where the $\phi^i$ are dynamical fields in the gravity side. These fields can be excited by turning on and off the charges of the system, thus, allowing the cosmological constant to vary. In this case, the natural quantity that we should identify with (\ref{IEgravity}) should be the Gibbs free energy, $G=U-TS-PV$ since, from the point of view of the extended thermodynamics, varying the cosmological constant is equivalent to varying the pressure $P=-\Lambda/8\pi G_N$; here $\Lambda=-3/L^2$ and $G_N$ is the Newton's constant.

The STU black holes considered here are solutions of gauged $N=8$, $D=4$ supergravity which arises as a truncation of $D=11$ supergravity compactified on a $S^{7}$. This theory can be uplifted to full M theory and hence, these black holes can be thought of as arising from different configurations of M2-branes. In the effective description, the value of $L$ is set by the value of the Planck length in the eleven dimensional theory $\lP$, and the number of branes $N$. The worldvolume theory is described in terms of a gauge theory with symmetries specified by the specific brane configuration; typically $N$ is the rank of the gauge group so it determines the number of degrees of freedom the theory. Moreover, the Newton's constant $G_N$ also depends on $N$. In four dimensions, these two quantities scale as $L\sim N^{1/6}$ and $G_N\sim N^{-7/6}$ \cite{Maldacena:1997re} so one finds that the pressure depends nontrivially in $N$, specifically as $P\sim N^{5/6}$. Is in this sense that, varying the cosmological constant $\Lambda$ (and hence the pressure), is equivalent to changing the field theory to which the bulk background is dual to.

One might wonder how this notion of RG-flow fits in the standard picture of the AdS/CFT correspondence. According to the UV/IR connection, the radial direction in the bulk maps to an energy scale in the boundary \cite{uvir1,uvir2,uvir3,uvir4}. This emergent direction can be thought of as way to encode the RG-flow in a geometrical way, since it links the UV and the IR of the theory (which correspond to the near-boundary and near-horizon portions of the geometry, respectively). However, in the framework of extended thermodynamics is the cosmological constant itself that plays the role of an energy scale, given that its value fixes the AdS radius $L$. We argue that the physical interpretation in each case is different. In order to understand better the distinction of these two scenarios let us review other relevant entries of the holographic dictionary. One of the main properties of the AdS/CFT correspondence is the map between bulk fields $\phi(x,r)$ and operators in the gauge
theory side $\cO(x)$. According to \cite{Gubser:1998bc,Witten:1998qj}, we can identify
\be\label{GKPWrecipe}
\mathcal{Z}[\phi_{0}] = \left\langle e^{-\int d^4x \phi_0(x) \mathcal{O}(x)} \right\rangle_{{\rm CFT}}\,,
\ee
where the term in the left hand side is the string partition function with boundary condition\footnote{Here assume the standard quantization.} $\phi\to r^{\Delta-3}\phi_0$ as $r\to\infty$ and the term in the right is the generating functional of correlation functions of the gauge theory. Notice that, for the right-hand side of (\ref{GKPWrecipe}) to be well-defined, $\cO$ has to have conformal weight $\Delta$, so that $\cO(\lambda x)=\lambda^{-\Delta}\cO(x)$. More in general, near the boundary, solutions to the equations of motion take the general form
\be
\phi\sim\frac{A}{r^{3-\Delta}}+\frac{B}{r^\Delta}+\cdots\,,
\ee
where $A$ and $B$ are identified with the source and expectation value of the dual operator $\cO$, respectively.
For the STU background, expanding the scalar fields (\ref{scalar1})-(\ref{scalar2}) around $r\to\infty$ lead to
\bea
&&\phi_{1} = \frac{q_{1}+q_{2}-q_{3}-q_{4}}{2r} + \frac{-q_{1}^{2}-q_{2}^{2}+q_{3}^{2}+q_{4}^{2}}{4r^{2}}+\cdots\,,\nonumber\\
&&\phi_{2} = \frac{q_{1}-q_{2}+q_{3}-q_{4}}{2r} + \frac{-q_{1}^{2}+q_{2}^{2}-q_{3}^{2}+q_{4}^{2}}{4r^{2}}+\cdots\,,\\
&&\phi_{3} = \frac{q_{1}-q_{2}-q_{3}+q_{4}}{2r} + \frac{-q_{1}^{2}+q_{2}^{2}+q_{3}^{2}-q_{4}^{2}}{4r^{2}}+\cdots\,.\nonumber\\
\eea
The conformal dimension of the three dual operators $\cO_i$, for $i=\{1,2,3\}$, is found to be $\Delta=2$.\footnote{In the alternative quantization the normalizable and non-normalizable modes interchange roles so $\Delta=1$. Either way $\Delta<3$ and the dual operator is relevant.} To see this, we can alternatively expand the potential (\ref{potdef}) as
\be\label{expanpot}
\mathcal{V} = -6 g^{2} - g^{2}(\phi_{1}^{2} + \phi_{2}^{2} + \phi_{3}^{2}) + \cO(\phi_i^4)\,,
\ee
from which we can read off the AdS radius $L=1/g$ and the three masses $m^2=-2g^2$. Then, we use the formula
\be\Delta=\frac{d+\sqrt{d^2+4m^2L^2}}{2}=2\,.
\ee
Thus, by giving different values to the charges in the bulk geometry we are, at the same time, turning on and off the couplings $\lambda_i$ of three \emph{relevant} operators in the gauge theory. Schematically, the Lagrangian of the dual theory is modified according to
\be
\mathcal{L}=\mathcal{L}_{0}+\sum_{i=1}^3\lambda_i\,\cO_i\,,
\ee
where $\mathcal{L}_{0}\sim g_{Y\!M}^2{\rm Tr}\,[F^2+\cdots]$ is the Lagrangian of the original theory (with all the charges/scalars turned off). Furthermore, regardless of the charges, the bulk geometry is asymptotically AdS, and thus the physics in the UV is still governed by the CFT prescribed by $\mathcal{L}_{0}$. However, the relevant operators $\cO_i$ can trigger an RG-flow to a new IR fixed point governed by a different theory. This effect is encoded in the radial dependence of the bulk geometry. Indeed, one of the effects of the non-trivial profile of the scalars is that the bulk solution is no longer invariant under the scale transformations $x^\mu\to\lambda x^\mu$, $r\to \lambda^{-1}r$ and therefore the solutions might have a completely different behavior deep in the bulk interior.

Now, we have argued that varying the cosmological constant provides an alternative notion of changing an energy scale of the dual theory. Can this notion be equivalent or, at the very least, related to the standard RG-flow encoded by the radial direction in the bulk? We believe that the answer is no. Rather, it provides a different and orthogonal direction in which we can move from theory to theory. Consider for example the AdS-RN black hole. In this case all the scalars are turned off so the Lagrangian of the theory is just $\mathcal{L}_{0}$, with no additional deformations. Here, it is clear that the non-trivial dependence on the radial coordinate just reflects the fact that we are not in the vacuum of the theory, but instead in an excited state with additional energy scales such as temperature and/or charge density. However, we can still think of varying the rank of the gauge group $N$, which can be achieved by changing the value of the cosmological constant $\Lambda$ (or, the pressure $P$). Hence, we can still have a non-trivial RG-flow. One issue that remains unclear in this context is how the volume $V$ should be interpreted in terms of the dual theory. In the case in which we have a non-trivial RG-flow along the radial direction, the relevant deformations $\cO_i$ break explicitly the conformal invariance and it is natural to define the theory at a cutoff scale $\Lambda_c$.\footnote{If all the $\lambda_i$ vanish as in the case of AdS-RN, conformal invariance is still broken due to the presence of temperature and/or charge density. In this scenario it is still sensible to define a cutoff scale $\Lambda_c$.} The dependence of physical quantities such as the couplings $\lambda_i(\Lambda_c)$ on the cutoff is encoded on the radial profile of the bulk solutions. Similarly, if we vary $N$ (or the pressure $P$) there should be an analogous scale $\Lambda_c^*$ that determines how the physical $N(\Lambda_c^*)$ runs along the RG-flow. This is precisely encoded in the $P(V)$ diagram at fixed $T$. Therefore, we can think of the volume $V$ as a cutoff scale that determines running of $N$ with energy.

Schematically, then, the $PV$ space can be thought of as a space theories with different $N$ and $\Lambda_c^*$; one parameter takes us from theory to theory while the other moves us along the RG-flow. We can, for instance, think  of  constructing ``thermodynamic'' cycles in this space of theories. These processes can be thought of as holographic heat engines, in the same sense of \cite{Johnson:2014yja}. In Appendix \ref{App} we show details of such a construction for the particular case of STU black holes.

Before closing this section let us comment on the nature of the phase transitions in the framework of the extended thermodynamics. We have seen that for the several cases studied, the relevant parameter that determines the transitions is the dimensionless parameter $\zeta=\beta^2P$. This provides us with an interesting possibility, namely, we can study the transitions by either by tuning the pressure (changing the theory) or the temperature (changing the state).
In terms of the $TS$ diagram at fixed $P$, the transitions occur to guarantee thermodynamic stability. In particular, the unstable branches with positive slope must be replaced by straight lines in order to satisfy the positivity of specific heat at constant pressure,
\be
C_P = T\left(\frac{\partial S}{\partial T}\right)_P\geq0\,.
\ee
Notice that this interpretation is valid for both, the bulk and the boundary theories. On the other hand, the analogous statement for the case of the $PV$ diagram at fixed $T$ is given in terms of the positivity of the isothermal compressibility\footnote{The minus sign in (\ref{kappadef}) accounts for the fact that an increase in pressure generally induces a reduction in volume.}
\be\label{kappadef}
\kappa_T=-\frac{1}{V}\left(\frac{\partial V}{\partial P}\right)_T\geq0\,,
\ee
where in this case the inequality stands for mechanical stability. The unstable branches in the $PV$ diagram should replaced by a straight line obeying the Maxwell's equal-area-law. However, since $P$ and $V$ are not related to the physical pressure and volume in the boundary theory, this interpretation does not hold in the dual QFT. Here, the analogous criterium for stability should be related to the monotonicity of $N(\Lambda_c^*)$ along the RG-flow! Exploring the consequences of this statement is clearly an interesting topic but lies beyond the scope of this paper. We will study it in more detail in the near future \cite{progress}.

\section{Holographic entanglement entropy}\label{Holographysec}
Having established that, depending on the charges, there are different kind of transitions in the extended phase space of STU black holes, we want to investigate how such transitions are reflected in the dual field theory. As mentioned in the introduction, the work of \cite{Johnson:2013dka} lead to the conclusion that at least for the AdS-RN case a similar transition to the one observed in the $TS$ diagram is present in the holographic computation of entanglement entropy. However, in that paper no connection with the extended thermodynamics was made. In this section, we will show that this transition is indeed the same as the transition appearing in the $PV$ space, implying that this field theory observable can be used as a tool to diagnose the extended phase structure of the theory. We will perform the calculation for various STU black holes and show that the same result holds more generally, thus making the conclusion more robust.

\subsection{Review of holographic entanglement entropy}\label{EEsubsec}

We will begin with a brief review of the Ryu-Takayanagi prescription to compute entanglement entropy. The formula was first conjectured in \cite{Ryu:2006bv,Ryu:2006ef} and later established rigorously in \cite{Lewkowycz:2013nqa}.

Let us start with some field theory generalities. At zero temperature, the system is in a pure state and can be described by a single state vector $| \Psi \rangle$, and by a density matrix $\rho = | \Psi \rangle \langle \Psi |$. The von Neumann entropy of the system is defined to be
\begin{equation}
S = -\mathrm{tr}\left[\rho \log{\rho}\right]\,.
\end{equation}
We can consider a Cauchy surface in spacetime, $\Sigma$, and let $A$ and $A^{c}$ be a subregion of $\Sigma$ and its complement, respectively. The Hilbert space factorizes into $\mathcal{H} = \mathcal{H}_{A} \otimes \mathcal{H}_{A^{c}}$, and one may define a reduced density matrix for region $A$ by tracing out the degrees of freedom of the complement,
\begin{equation}
\rho_{A} = \mathrm{tr}_{A^{c}}[\rho]\,.
\end{equation}
The entanglement entropy of region $A$ is then defined as the von Neumann entropy of the reduced density matrix,
\begin{equation}
S_{A} = -\mathrm{tr}[\rho_{A}\log{\rho_{A}}]\,.
\end{equation}
Consider now the finite temperature case. For a local quantum field theory, the system's Hamiltonian $H$ can be split into $H = H_{A} + H_{A^{c}} + H_{\partial A}$, where $H_{A}$ is the Hamiltonian of the degrees of freedom in the interior of $A$, $H_{A^{c}}$ is the Hamiltonian of the complement of $A$ and $H_{\partial A}$ is the Hamiltonian of the interface of $A$ with its complement. The total density matrix is then given by:
\begin{equation}
\rho = e^{-\beta H}\,,
\end{equation}
and the definitions of the von Neumann entropy and entanglement entropy are the same as in the zero temperature case.

Let us now move to the holographic prescription. According to the Ryu-Takayanagi formula, $S_A$ is computed \emph{a la} Bekenstein-Hawking,
\begin{equation}
S_{A} = \frac{\text{Area}(\Gamma_A)}{4 G_N}\,,
\end{equation}
where $\Gamma_A$ is a codimension-2 minimal surface with boundary condition $\partial \Gamma_A=\partial A$, and $G_{N}$ is the gravitational Newton's constant.\footnote{The notion of a minimal surface only makes sense in Euclidean space. Therefore, the formula stated above only works for static spacetimes, where, by symmetry, the minimal surface lies on a constant time slice. For time-dependent backgrounds, the prescription can be made covariant by replacing the minimal surface condition by an extremal surface condition \cite{Hubeny:2007xt}.} 
Notice the close similarity with equation (\ref{bhent}). Indeed, if we consider $A$ to approach the full $\Sigma$, it is possible to show that part of $\Gamma_A$ tends to wrap the horizon of the black hole and the leading term of the entanglement entropy is dominated by thermal entropy, as expected.

Finally, note that in a compact space, the complement of $A$ basically has the same shape as $A$ so one might naively think that the same minimal surface computes both $S_{A}$ and $S_{A^{c}}$. This is indeed true in the vacuum because in any pure state $S_{A}=S_{A^{c}}$. However, in a mixed state this is no longer true. In this case there is an additional topological requirement for the minimal surface in the bulk \cite{Haehl:2014zoa}: $\Gamma_{A}$ has to be homologous to $A$. If there is a horizon in the bulk, the minimal surface $\Gamma_A$ is homologous to $A$ but not to $A^{c}$. To satisfy the homology requirement for $S_{A^{c}}$, we have to include a second component: the surface of the horizon itself. For more details on the homology requirement, see for example \cite{Hubeny:2013gta,toappear}.

\subsection{Entanglement entropy across the phase transitions}

In this subsection, we track entanglement entropy across the various phase transitions of the STU black hole and we will argue that entanglement entropy detects the Van der Waals transition whenever it occurs on the gravity side.\footnote{See \cite{Albash:2012pd,Kol:2014nqa,Georgiou:2015pia} for related works on holographic entanglement entropy and phase transitions.}

Consider a constant time slice, and take the region $A$ to be a spherical cap on the boundary delimited by $\theta \leq \theta_{0}$. We will pick a small value for $\theta_{0}$ because we want to filter out the thermal part of the entropy. Specifically, we will choose $\theta_{0} = 0.15$. Now, since entanglement entropy is UV-divergent (as is the area of the minimal surface), it needs to be regularized. There are several ways to accomplish this; we will do it by subtracting the area of the minimal surface in pure AdS. In order to do so, we first integrate the area functional to some cutoff $\theta_{c} \lesssim \theta_{0}$. Then we turn off the mass and all the charges of the black hole to obtain pure AdS in global coordinates:
\begin{equation}
ds^{2} = -(1+g^{2}r^{2})dt^{2} + \frac{dr^{2}}{1+g^{2}r^{2}} + r^{2}d\Omega^{2}\,.
\end{equation}
The minimal surface which goes to $\theta=\theta_{0}$ on the boundary is given by
\begin{equation}
r{(\theta)} = \frac{1}{g}\left(\left(\frac{\cos{\theta}}{\cos{\theta_{0}}}\right)^{2}-1\right)^{-1/2}\,.
\end{equation}
We can easily integrate the area functional of this minimal surface up to $\theta_{c}$. Then we subtract this quantity from the black hole one, and send $\theta_{c}$ to $\theta_{0}$. This gives us the renormalized entanglement entropy, which we will denote by $\Delta S_{A}$. In particular, for the numerical computations we will choose $\theta_{c}$ to be $0.149$.

\begin{figure*}[t!]
$$
\begin{array}{cc}
 \includegraphics[width=8cm]{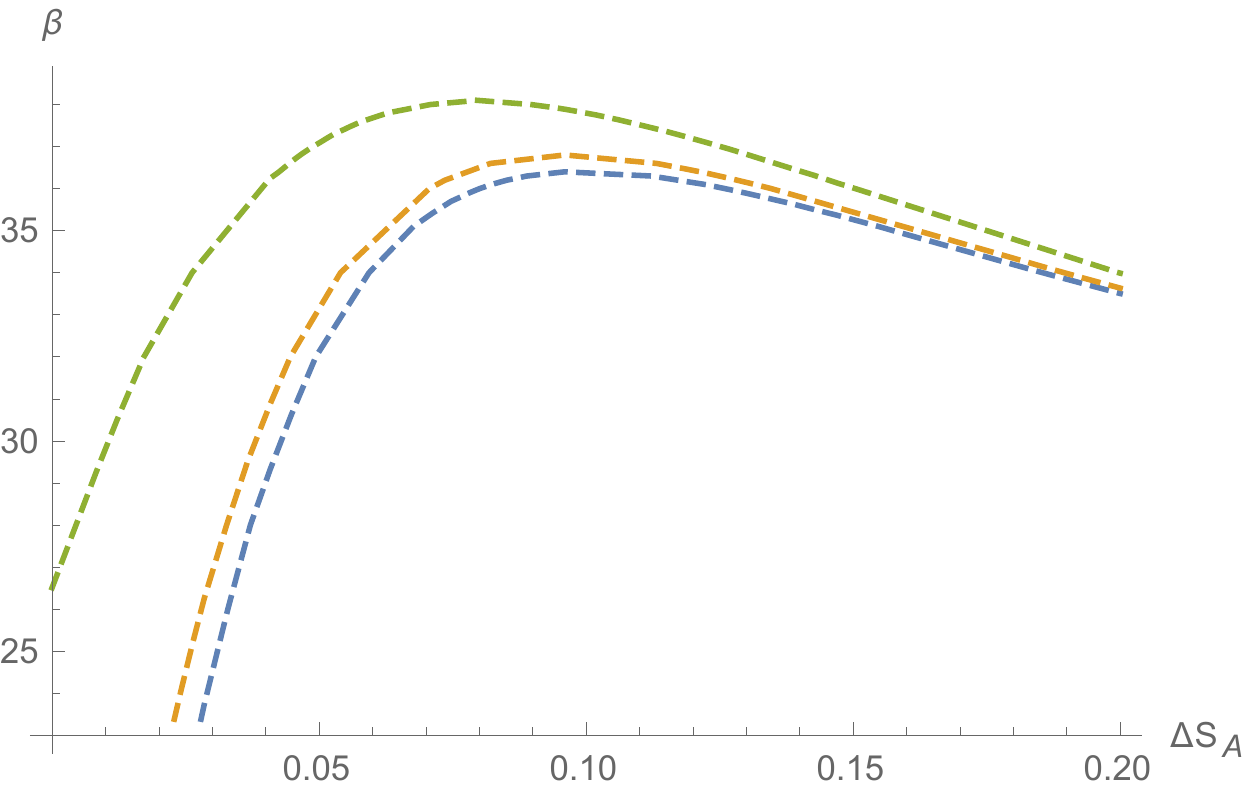} & \includegraphics[width=8cm]{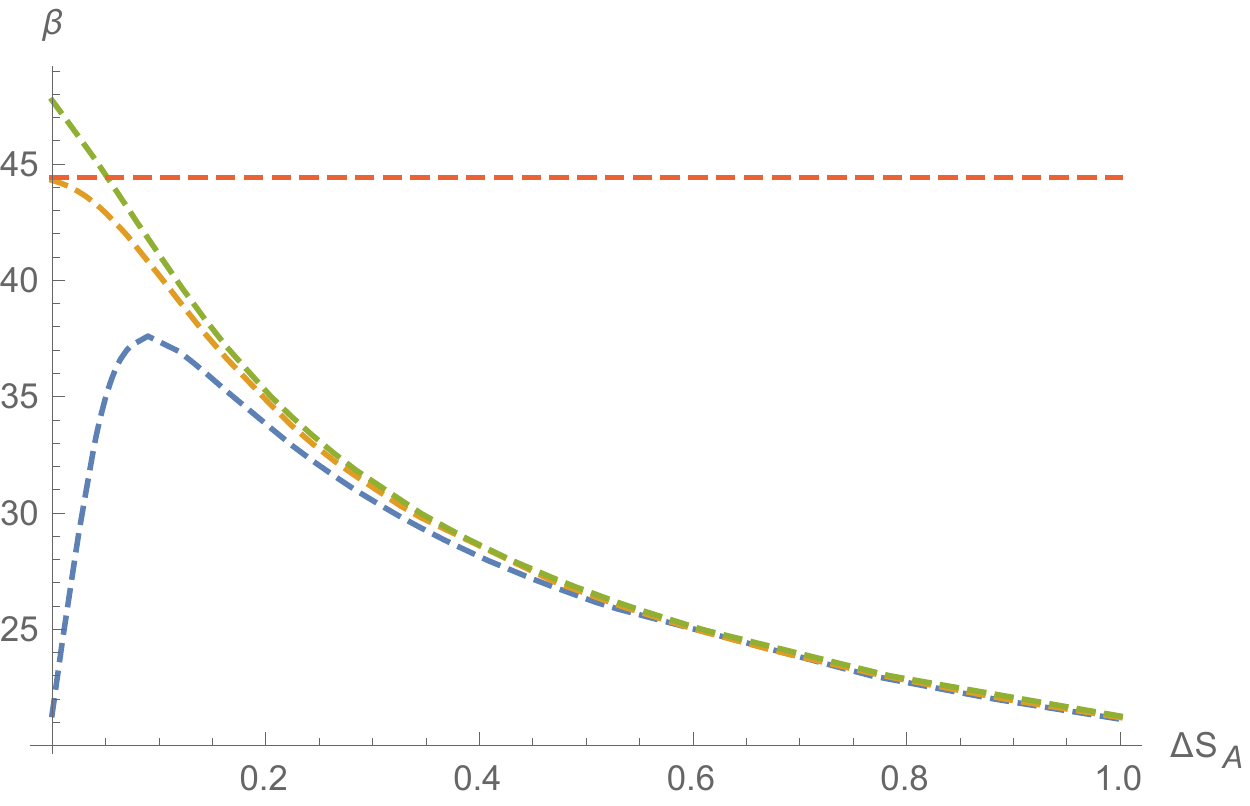}\\
 \quad\quad(a) & \quad\quad(b)\\
  \includegraphics[width=8cm]{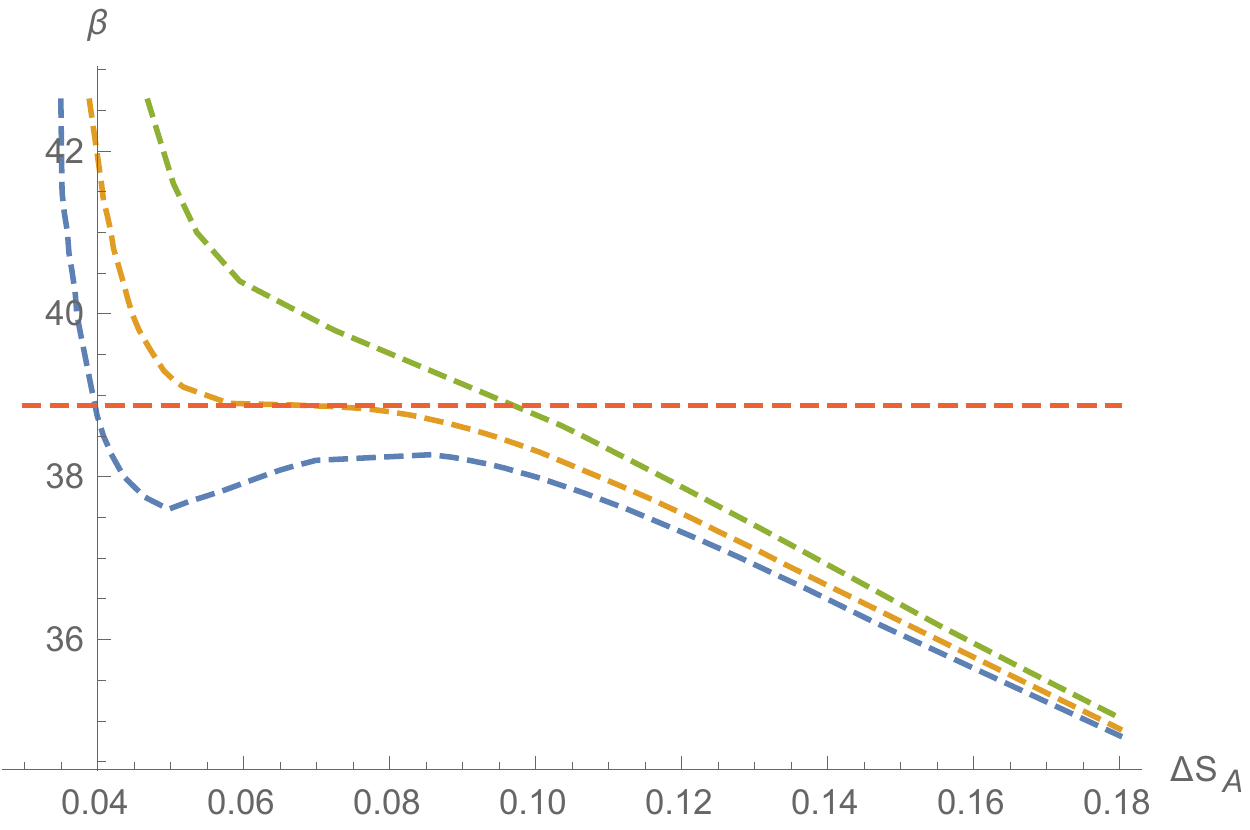} & \includegraphics[width=8cm]{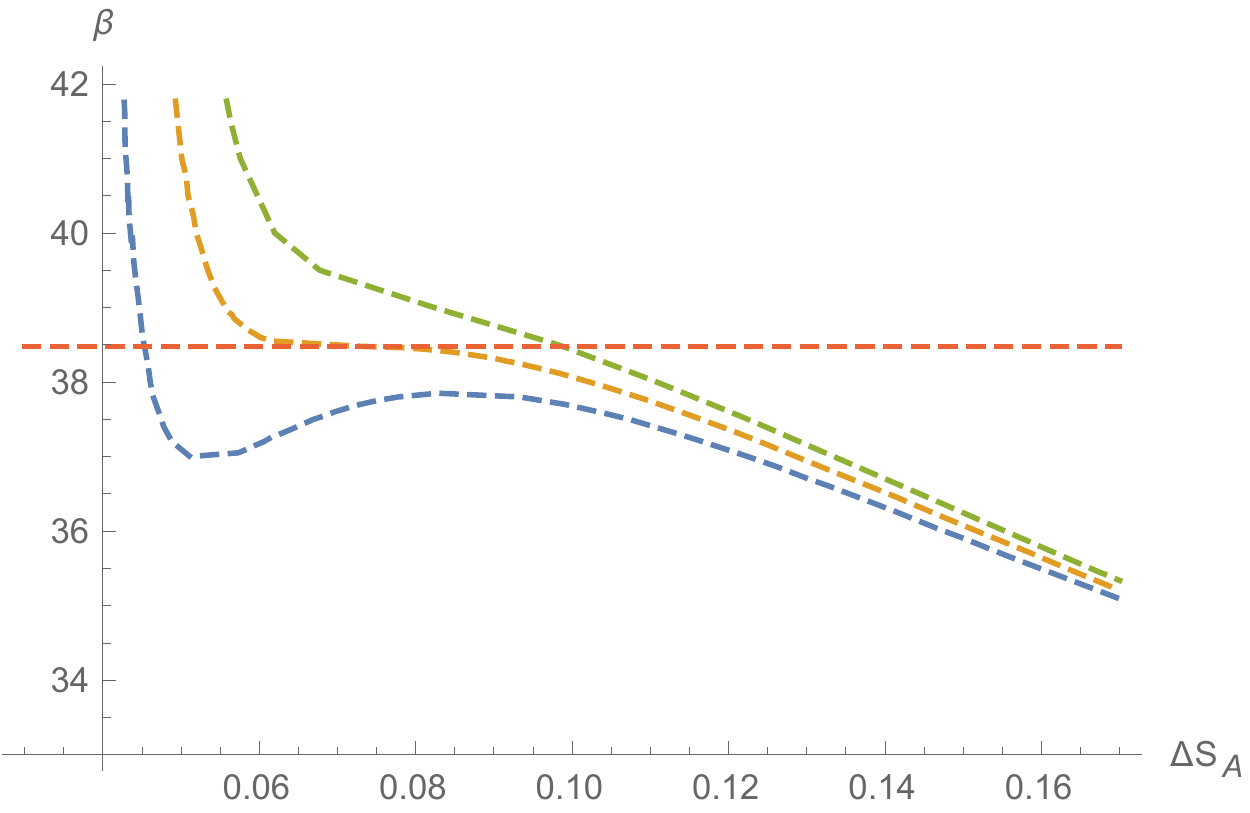}\\
 \quad\quad(c) & \quad\quad(d)\\
\end{array}
$$
\vspace{-0.7cm}
\caption{Plots of entanglement entropy versus $\beta$.
Panel (a): The single-charge case, with $Q=2$ (green), $Q=1$ (orange) and $Q=0.5$ (blue). Panel (b): The two-equal-charge case, with $Q = 2.5 > Q_{c}$ (green), $Q = 2.0704 = Q_{c}$ (orange) and $Q = 1 < Q_{c}$ (blue). The critical temperature is $\beta_{c} = 44.4187$.
Panel (c): The three-equal-charge case, with $Q = 1.5 > Q_{c}$ (green), $Q = 1.0478 = Q_{c}$ (orange), and $Q = 0.97 < Q_{c}$ (blue). The critical temperature is $\beta_{c} = 38.8714$.
Panel (d): the AdS-RN case with $Q = 0.9 > Q_{c}$ (green), $Q = Q_{c} = 5/6$ (orange) and to $Q = 0.75 < Q_{c}$ (blue). The critical temperature is $\beta_{c} = 10\pi\sqrt{3/2}$.}
\label{EEvsbetafig}
\end{figure*}

The plots generated for the entanglement entropy are shown in Figure \ref{EEvsbetafig}.
Let us start the discussion with the AdS-RN case, which was treated in \cite{Johnson:2013dka} but is included here for comparison's sake. In Figure \ref{EEvsbetafig}(d), we plot $\Delta S_{A}$ versus $\beta$ for three different values of the charge (above, at, and below the critical charge) and confirm that there is indeed a Van der Waals transition. Moreover, the critical charge and critical temperature agree with those for the $TS$ transition ---see Figure \ref{betavsSfig}--- and with the formulas (\ref{ScRNAdS}). Since we had already established that the $TS$ transition captures the same physical information as the $PV$ transition, we can conclude that the entanglement entropy also has access to the extended phase structure. For large enough regions this is indeed expected, since entanglement entropy is dominated by thermal entropy. However, the non-trivial result we point out here is that this behavior holds true even for arbitrary small regions. Indeed, we find that the critical charge and temperature are independent of the size $\theta_{0}$ of the boundary disk.

The results for the three-equal-charge case are shown in Figure \ref{EEvsbetafig}(c). As in the AdS-RN case, the three-charge case also presents a Van der Waals transition, and the $\Delta S_{A}$ versus $\beta$ plot is qualitatively similar to Figure \ref{betavsSfig}(c). The critical temperature and critical charge also agree with the ones obtained from Figure \ref{betavsSfig}, providing further evidence on our previous conclusion. Unlike the two previous cases, the two-equal-charge case and the single-charge case do not exhibit the Van der Waals transition, again like in Figure \ref{betavsSfig}. These plots are shown in
\ref{EEvsbetafig}(b) and \ref{EEvsbetafig}(a), respectively. For the two-equal-charge case there is an unstable branch at subcritical charge, in analogy with Figure \ref{betavsSfig}(b); this branch is squeezed out when the critical charge and the critical temperature are attained, with values that agree with those in Figure \ref{betavsSfig}. Finally, the single-charge case does not have a Van der Waals transition, like in Figure \ref{betavsSfig}.

\subsection{Critical exponent of entanglement entropy\label{cexpee}}
Given that the entanglement entropy behaves very similarly to the black hole entropy, even for away from the thermal regime, it is natural to wonder if it is possible to define a ``specific heat'' at constant $P$ for entanglement entropy:
\begin{equation}
\tilde{C}_{P} = T\left(\frac{\partial \Delta S_{A}}{\partial T}\right)_{P}\,,
\end{equation}
as well as an associated critical exponent $\alpha$. For the three-equal-charge case, we have plotted $\log{|\beta-\beta_{c}|}$ versus $\log|\Delta S_{A}-\Delta S_{A,c}|$. To produce this plot, we used the critical curve of Figure \ref{EEvsbetafig}(c), and in particular the 32 points which are closest to the inflection point on this curve.
\begin{figure*}[t!]
$$
 \includegraphics[width=10cm]{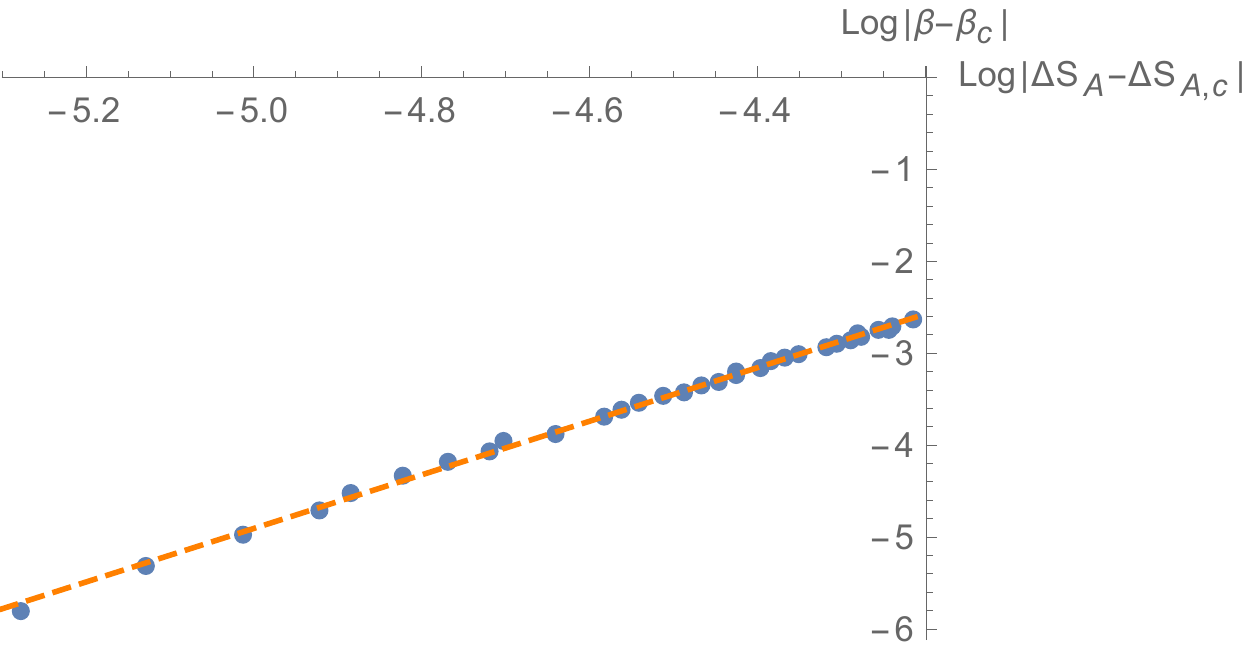}
$$
\vspace{-0.5cm}
\caption{Plot of $\log{|\beta-\beta_{c}|}$ versus $\log|\Delta S_{A}-\Delta S_{A,c}|$ based on the 32 points closest to the inflection point of Figure \ref{EEvsbetafig} (c), ranging from $\Delta S_{A}=0.065$ (the critical value) to $\Delta S_{A}=0.08$.}
\label{EEcritexpfig}
\end{figure*}
A linear fit is shown in Figure \ref{EEcritexpfig} and yields:
\begin{equation}
\log{|\beta-\beta_{c}|} = 2.91447\log{|\Delta S_{A} - \Delta S_{A,c}|} + 9.665
\end{equation}
The slope is consistent with the value $3$, and therefore the critical exponent $\alpha$ for entanglement entropy agrees with that of the black hole entropy. In particular, it seems that this critical exponent is independent of the size $\theta_{0}$ of the entangling region. However, there could potentially be a weak dependence on $\theta_{0}$ that can only be detected at higher numerical accuracy. We will leave this interesting issue for future work.

\section{Conclusions}\label{Conclusion}

The extension of black hole thermodynamics by uplifting the cosmological constant to a dynamical variable is a fascinating topic that has generated a lot of recent interest. According to this framework, black holes seem to undergo critical phenomena very similar to ordinary matter. For instance, the extended phase structure of the AdS-RN black hole reinforces the well-known fact that this system exhibits a Van der Waals-like liquid-gas transition. Similarly, other AdS black holes are shown to exhibit a very rich structure that includes a whole catalogue of classical phase transitions, such as reentrant phase transitions and multicritical points, making them a useful tool for holographic applications of condensed matter systems.

In this paper, we have investigated the extended thermodynamics of the 4-dimensional STU black hole in the fixed charge ensemble. We uncovered an interesting $PV$ phase structure that interpolates between the Hawking-Page transition of the Schwarzschild-AdS solution and the Van der Waals transition of AdS-RN, depending of the number of charges turned on. When only one charge is present, we still have a Hawking-Page-like transition. When two charges are turned on, we have an intermediate phase structure that resembles the Hawking-Page transition at sufficiently low temperatures. Finally, when three or four charges are turned on, we have a Van der Waals-like transition. One might wonder if the nature of these two transitions is the same, since they both have a qualitatively similar behavior. In order to answer this question we have extracted numerically two of the critical exponents that characterize the transition for the three-charge case, and confirmed that they are indeed the same as those of the AdS-RN black hole. Therefore, the three-charge and AdS-RN black holes seem to belong to the same universality class, although it would be interesting to extend our analysis to all the other critical exponents. We will leave this interesting question for future work.

We also showed that the transitions appearing in the $PV$ space are also present in the $TS$ diagrams and explained it by arguing that the relevant parameter is always a dimensionless combination between the temperature and the pressure, $\zeta=\beta^2P$. Thus, we can approach the transition either by varying $P$ or $T$ and keeping the other quantity fixed. The interesting point here is that, in terms of the dual field theory, these two process have completely different physical meaning; in one case one is changing the field theory while in the other case one is changing the state. This is due to the fact that by varying the cosmological constant we are changing the rank of the gauge group of the boundary theory, $N$, and hence the $PV$ diagram can be interpreted as a space of theories. We argued that the thermodynamical volume $V$ has a natural interpretation as a cutoff scale $\Lambda_c^*$, so that the isotherms in a $PV$ diagram
are interpreted as the running of $N(\Lambda_c^*)$ with the energy. In this context, then, the phase transitions are interpreted as a requirement for the monotonicity of $N(\Lambda_c^*)$ along the RG-flow.

Now, given that thermal entropy is subject to the same phase transitions of the extended $PV$ phase space, it is natural to wonder if the same phenomenon can be observed for other field theory observables. One natural possibility to consider is entanglement entropy, which is known to reduce to thermal entropy for large enough regions. Indeed, previous studies have shown that entanglement entropy also exhibits a Van der Waals-like transition for the AdS-RN black hole,
but no connection with the extended thermodynamics was pointed out. Remarkably, we have shown that even for entangling regions of small size, the Van der Waals transition is still present, and it happens at exactly the same critical temperature, pressure and charge. This shows that entanglement entropy have also access to the extended thermodynamics and can be used as an efficient tool to diagnose its phase structure. In addition, the same statement seems to hold for other charge configurations of the STU black hole, regardless the nature of the transition. This provides further evidence supporting our results and makes the conclusion more robust. Finally, we computed numerically the critical exponent for the entanglement entropy and showed that it coincides with that of the thermal entropy whenever the Van der Waals transition is present.

A number of possibilities for the extension of our work can be considered. First, we can perform a similar study in backgrounds that exhibit peculiar properties. One of such examples is the AdS-Taub-NUT geometry, which is known to have negative thermodynamical volume \cite{Johnson:2014xza,Johnson:2014pwa,Lee:2014tma}. It would be interesting to see how this case would fit in the interpretation presented in this paper. Similarly, there are other examples that belong to this category. Two of them are the AdS-Kerr-bolt black hole and the Lifshitz black hole \cite{MacDonald:2014zaa,Brenna:2015pqa}. In the first case, the issue arises because the volume is not the same as the one arising from naive geometrical considerations while, in the second, there is an ambiguity in expressing the cosmological constant lengthscale in terms of the horizon radius lengthscale. We can also consider the extended thermodynamics in dynamical settings, \emph{e.g.} in Vaidya backgrounds, at least in the adiabatic limit. In such a regime one can still define an effective thermodynamics in terms of an apparent horizon and one might encounter interesting phenomena, specially in charged backgrounds \cite{Galante:2012pv,Caceres:2012em,Caceres:2013dma,Caceres:2014pda}. A second line of study would be to investigate whether these phase transitions can be observed or not in other field theory observables. Some natural possibilities to explore are Wilson loops, correlation functions, or other entanglement-related quantities such as mutual information or renormalized entanglement entropy (a good starting point would be \cite{Bhattacharyya:2014oha}). Finally, it would also be interesting to explore how $1/N$ corrections affect the phase structure of the extended thermodynamics and the entanglement entropy. Since our results are strictly valid in the large-$N$ limit, one might wonder if the phase transitions we observed here can receive corrections, or even disappear (see \emph{e.g.} \cite{Dolan:2010ha}).




\section*{Acknowledgements}
It is a pleasure to thank Willy Fischler, Clifford Johnson, Rex Lundgren and Carlos Nu\~nez for discussions and comments on the manuscript. E.C  thanks the Galileo Galilei Institute for Theoretical Physics and the KITP  for  hospitality and the INFN for partial support during the completion of this work.
This research was supported by the National Science Foundation under Grant PHY-1316033 and Grant NSF PHY11-25915, and by Mexico's National Council of Science and Technology
(CONACyT) grant CB-2014-01-238734 (EC).

\appendix

\section{Holographic heat engines}\label{App}

Treating the cosmological constant as a dynamical variable grants us access to the $PV$ phase space which, as explained in Section \ref{Interpretation}, can be understood as a space of field theories with different $N$ and cutoff scale $\Lambda_c^*$. Thus, we can think of constructing several cycles by moving in this RG-flow along different paths. Such processes can be thought of as holographic ``heat engines'', in the sense of \cite{Johnson:2014yja}.
In this section, we present a few analytic results about STU black holes as heat engines, especially the efficiency of these systems in the high pressure regime. The results presented here are an extension of the paper \cite{Johnson:2014yja}, where the efficiency of the AdS-RN case was studied.\footnote{Other holographic heat engines were constructed recently in \cite{Sadeghi:2015ksa}.}

For any cycle, the efficiency $\eta$ is defined to be:
\begin{equation}
\eta = \frac{W}{Q_{H}}\,,
\end{equation}
where $W$ is the work generated during one cycle, and $Q_{H}$ is the net heat flowing into the engine. Since $W = Q_{H}-Q_{C}$, where $Q_{C}$ is the waste heat leaving the engine, the efficiency can be equivalently defined as:
\begin{equation}
\eta = 1 - \frac{Q_{H}}{Q_{C}}\,.
\end{equation}
In a reversible process, the entropy leaving the engine must equal the entropy flowing into the engine. Since $Q = TS$, this implies
\begin{equation}
Q_{C} = \frac{T_{C}}{T_{H}}Q_{H}\,,
\end{equation}
and therefore the efficiency of a reversible process takes the form
\begin{equation}
\eta = \eta_{C} \equiv 1 - \frac{T_{C}}{T_{H}}\,.
\end{equation}
This efficiency is called the Carnot efficiency, and is the maximal value that any heat engine can achieve. Here, we will verify that none of the STU black holes can violate the Carnot bound for the efficiency.

Let us focus on the large-$P$ limit or, equivalently, large-$N$ limit. For the single-charge case, the high pressure expansions of the temperature, entropy and volume are (working to second order in $g^{-2}$):
\begin{equation}\label{eq:ThighP1charge}
T = \frac{3}{4\pi}r_{+}g^{2} + \frac{2 Q^{2}+r_{+}^{2}}{4\pi r_{+}^{3}} - \frac{3Q^{4}+2Q^{2}r_{+}^{2}}{2\pi g^{2}r_{+}^{7}} + \cdots\,,
\end{equation}
\begin{equation}\label{eq:ShighP1charge}
S = \pi r_{+}^{2} + \frac{2\pi Q^{2}}{r_{+}^{2}g^{2}} - \frac{1}{g^4}\left(\frac{2\pi Q^{2}}{r_{+}^{4}}+\frac{10\pi Q^{4}}{r_{+}^{6}}\right) + \cdots\,,
\end{equation}
\begin{equation}\label{eq:VhighP1charge}
V = \frac{4}{3}\pi r_{+}^{3} + \frac{4\pi Q^{2}}{r_{+}g^{2}} -\frac{1}{g^{4}}\left(\frac{4\pi Q^{2}}{r_{+}^{3}}+\frac{16\pi Q^{4}}{r_{+}^{5}}\right) + \cdots\,.
\end{equation}
For the two-equal-charge black hole, the expansions are:
\begin{equation}\label{eq:ThighP2charge}
T = \frac{3}{4\pi}r_{+}g^{2} + \frac{4 Q^{2}+r_{+}^{2}}{4\pi r_{+}^{3}} - \frac{2(4Q^{4}+Q^{2}r_{+}^{2})}{\pi g^{2}r_{+}^{7}} + \cdots\,,
\end{equation}
\begin{equation}\label{eq:ShighP2charge}
S = \pi r_{+}^{2} + \frac{4\pi Q^{2}}{r_{+}^{2}g^{2}} - \frac{1}{g^4}\left(\frac{4\pi Q^{2}}{r_{+}^{4}}+\frac{32\pi Q^{4}}{r_{+}^{6}}\right) + \cdots\,,
\end{equation}
\begin{equation}\label{eq:VhighP2charge}
V = \frac{4}{3}\pi r_{+}^{3} + \frac{8\pi Q^{2}}{r_{+}g^{2}} -\frac{1}{g^{4}}\left(\frac{8\pi Q^{2}}{r_{+}^{3}}+\frac{160\pi Q^{4}}{3r_{+}^{5}}\right) + \cdots\,.
\end{equation}
Finally, for the three-equal-charge black hole, the expansions are:
\begin{equation}\label{eq:ThighP3charge}
T = \frac{3}{4\pi}r_{+}g^{2} + \frac{6 Q^{2}+r_{+}^{2}}{4\pi r_{+}^{3}} - \frac{3(13Q^{4}+2Q^{2}r_{+}^{2})}{2\pi g^{2}r_{+}^{7}} + \cdots\,,
\end{equation}
\begin{equation}\label{eq:ShighP3charge}
S = \pi r_{+}^{2} + \frac{6\pi Q^{2}}{r_{+}^{2}g^{2}} - \frac{1}{g^4}\left(\frac{6\pi Q^{2}}{r_{+}^{4}}+\frac{66\pi Q^{4}}{r_{+}^{6}}\right) + \cdots\,,
\end{equation}
\begin{equation}\label{eq:VhighP3charge}
V = \frac{4}{3}\pi r_{+}^{3} + \frac{12\pi Q^{2}}{r_{+}g^{2}} -\frac{1}{g^{4}}\left(\frac{12\pi Q^{2}}{r_{+}^{3}}+\frac{112\pi Q^{4}}{r_{+}^{5}}\right) + \cdots\,.
\end{equation}
If we keep only the leading term in each of the expansions above, then we obtain an approximate equation of state valid in the high pressure, high temperature regime:
\begin{equation}\label{eoshighP}
P \approx \frac{1}{2}\left(\frac{4\pi}{3}\right)^{1/3}\frac{T}{V^{1/3}}\,.
\end{equation}
Comparing with the equation of state of the AdS-RN solution, we find that the right-hand side of (\ref{eoshighP}) is the first term in (\ref{RNAdSeos}). Therefore, all four cases of the STU black hole have the same equation of state in the high pressure, high temperature limit. However, due to the subleading terms, the entropy and volume in general no longer determine each other, unlike the AdS-RN case. For the single-charge case we have:
\begin{equation}\label{eq:SVhighP1charge}
V = \frac{4}{3\sqrt{\pi}}S^{3/2} + \left(\frac{3\pi^{2}}{32}\right)^{1/3}\frac{Q^{4}}{P^{2}V^{5/3}} + \cdots\,,
\end{equation}
for the two-equal-charge case:
\begin{equation}\label{eq:SVhighP2charges}
V = \frac{4}{3\sqrt{\pi}}S^{3/2} + \left(\frac{2\pi^{2}}{9}\right)^{1/3}\frac{Q^{4}}{P^{2}V^{5/3}} + \cdots\,,
\end{equation}
and for the three-equal-charge case:
\begin{equation}\label{eq:SVhighP3charges}
V = \frac{4}{3\sqrt{\pi}}S^{3/2} + \left(\frac{3\pi^{2}}{32}\right)^{1/3}\frac{Q^{4}}{P^{2}V^{5/3}} + \cdots\,.
\end{equation}
Notice that the leading term on the right-hand side of (\ref{eq:SVhighP1charge})-(\ref{eq:SVhighP3charges}) is exactly the Reissner-Nordstrom relation (\ref{eq:SVAdSRN}). The second terms have a common form but they differ by a multiplicative numerical constant.

\subsection{Isotherm-isochore cycle}
\begin{figure*}[t!]
\center
 \includegraphics[width=8cm]{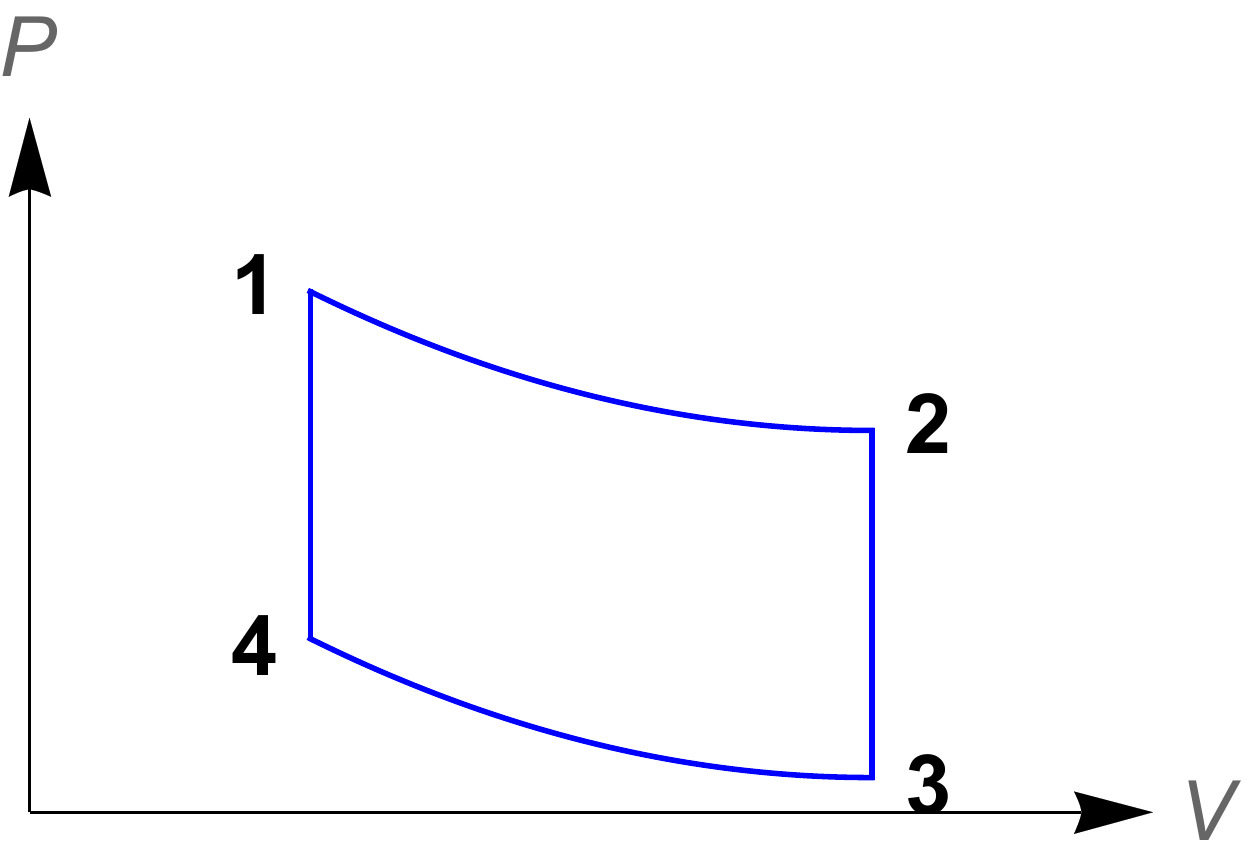}
\caption{The isotherm-isochore engine. The curvy lines are isotherms, and the straight lines are isochores. In terms of the boundary theory, an isochore can be interpreted as a process where $N$ is varied while keeping the cutoff $\Lambda_c^*$ fixed.}
\label{Stirlingenginefig}
\end{figure*}

In this subsection, we compute the efficiency of Stirling cycles, \emph{i.e.} cycles consisting of two isotherms and two isochores ---see Figure \ref{Stirlingenginefig}. In the case of the AdS-RN, the fact that $S$ and $V$ are proportional to each other implies that the adiabatic curves coincide with the isochoric ones, and Stirling cycles coincide with Carnot cycles. Therefore, the Stirling cycle for AdS-RN operates at the optimal efficiency $\eta_{C}$.
To see this, consider the work $Q_{H}$, $Q_{C}$ done along the upper and lower isotherms:
\begin{equation}\label{QHRN}
Q_{H} = T_{H}\Delta S_{1\rightarrow 2} = T_{H} \left(\frac{3\sqrt{\pi}}{4}\right)^{2/3}(V_{2}^{2/3}-V_{1}^{2/3})\,,
\end{equation}
\begin{equation}\label{QCRN}
Q_{C} = T_{C}\Delta S_{3\rightarrow 4} = T_{C} \left(\frac{3\sqrt{\pi}}{4}\right)^{2/3}(V_{3}^{2/3}-V_{4}^{2/3})\,,
\end{equation}
where the subscripts $1$, $2$, $3$, $4$ refer to the points labeled on Figure \ref{Stirlingenginefig}. Since $V_{1}=V_{4}$ and $V_{2}=V_{3}$, we deduce that the Carnot efficiency is attained.

Next, consider the single-charge black hole. Using the expansion (\ref{eq:SVhighP1charge}), we now find that the heat flows along the isotherms pick up some corrections:
\begin{equation}
Q_{H} = Q_{H}^{\text{RN}} + T_{H}\frac{\pi Q^{4}}{4}\left(\frac{1}{V_{1}^{2}P_{1}^{2}}-\frac{1}{V_{2}^{2}P_{2}^{2}}\right)\,,
\end{equation}
\begin{equation}
Q_{C} = Q_{C}^{\text{RN}} + T_{C}\frac{\pi Q^{4}}{4}\left(\frac{1}{V_{4}^{2}P_{4}^{2}}-\frac{1}{V_{3}^{2}P_{3}^{2}}\right)\,,
\end{equation}
where $Q_{H}^{\text{RN}}$ and $Q_{C}^{\text{RN}}$ are given by (\ref{QHRN}) and (\ref{QCRN}), respectively.
Next, we use the high pressure equation of state (\ref{eoshighP}) to rewrite the correction terms above:
\begin{equation}
Q_{H} = Q_{H}^{\text{RN}} + \frac{Q^{4}}{T_{H}}\left(\frac{9\pi}{16}\right)^{1/3}\left(\frac{1}{V_{1}^{4/3}}-\frac{1}{V_{2}^{4/3}}\right)\,,
\end{equation}
\begin{equation}
Q_{C} = Q_{C}^{\text{RN}} + \frac{Q^{4}}{T_{C}}\left(\frac{9\pi}{16}\right)^{1/3}\left(\frac{1}{V_{4}^{4/3}}-\frac{1}{V_{3}^{4/3}}\right)\,.
\end{equation}
It follows that the efficiency is:
\begin{equation}
\eta = 1 - \frac{T_{C}}{T_{H}}\left(1+Q^{4}\frac{V_{1}^{2/3}+V_{2}^{2/3}}{(V_{1}V_{2})^{4/3}}\left(\frac{1}{T_{C}^{2}}-\frac{1}{T_{H}^{2}}\right)\right)\,,
\end{equation}
where we have used the fact that $V_{1} = V_{4}$ and $V_{2}=V_{3}$. Since $T_{C} < T_{H}$, it follows from the above that the efficiency is less than the Carnot efficiency. The efficiency for the two-charge and three-charge cases can be obtained from the above by rescaling the common charge $Q$ and its found to be less than the Carnot efficiency as well.

\subsection{Isobar-isochore cycle}
\begin{figure*}[t!]
\center
 \includegraphics[width=8cm]{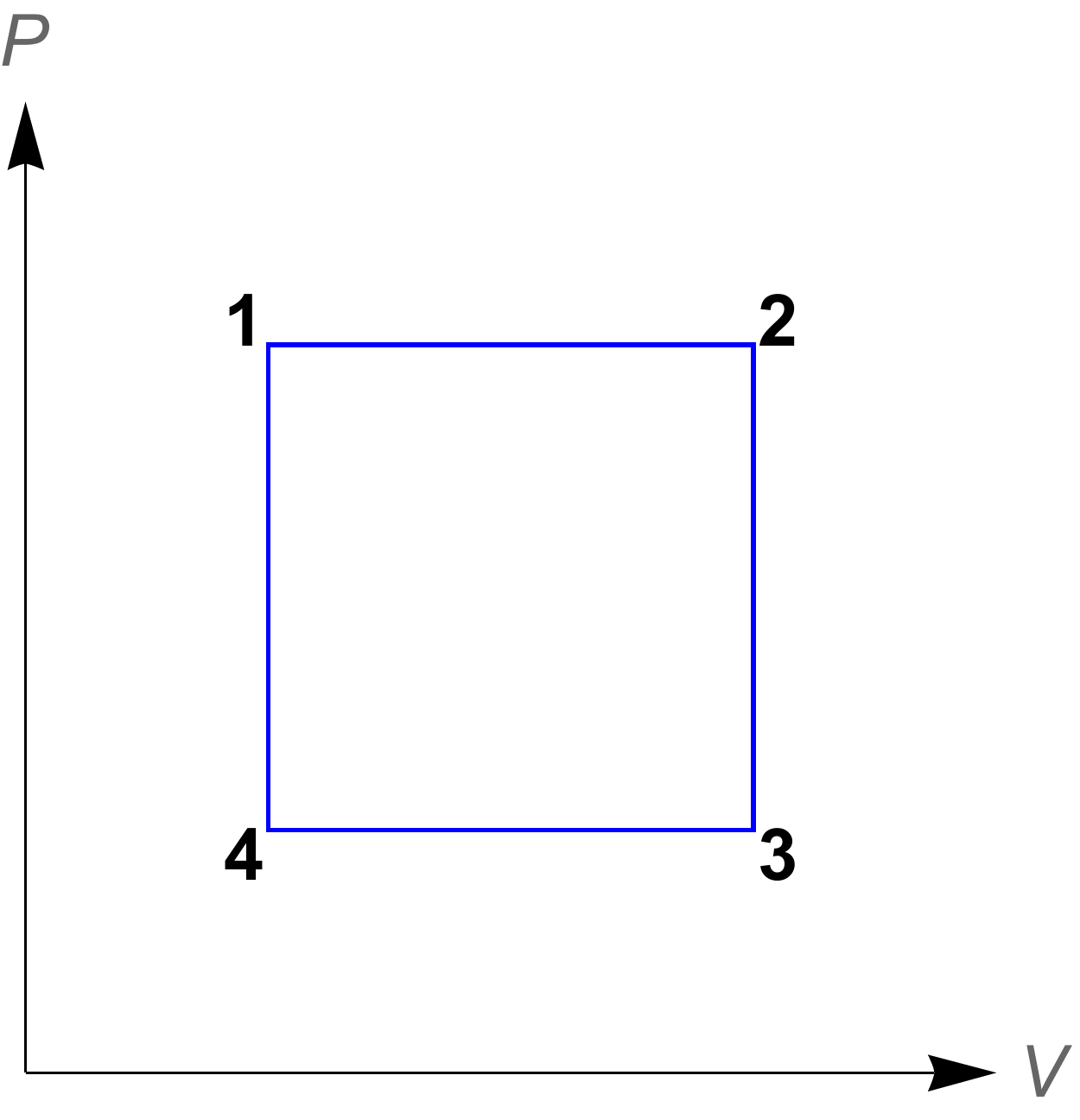}
\caption{The isobar-isochore engine. The vertical lines are isochores, and the horizontal lines are isobars. In terms of the boundary theory, an isochore can be interpreted as a process where $N$ is varied while keeping the cutoff $\Lambda_c^*$ fixed. In an isobar process, on the other hand, $\Lambda_c^*$ is varied while $N$ is kept fixed.}
\label{secondenginefig}
\end{figure*}
Next, consider a cycle composed of a rectangle on the $PV$ diagram, \emph{i.e.} two isochores and two isobars, as depicted in Figure \ref{secondenginefig}. For this cycle, it is enough to truncate the high temperature expansions to first order in $g^{-2}$. To this order, all the three special cases (one-, two- and three-equal-charges) can be treated in the same time by defining an effective charge $Q_{eff}$ as follows:
\begin{equation}
Q_{eff}^{2} = Q_{1}^{2}+Q_{2}^{2}+Q_{3}^{2}+Q_{4}^{2}\,.
\end{equation}
In particular, $Q_{eff} = Q$ for the single-charge case, $Q_{eff} = \sqrt{2}Q$ for the two-equal-charge case, and $Q_{eff}=\sqrt{3}Q$ for the three-equal-charge case. To compute the heat flow along the isobars, we need the specific heat at constant pressure:
\begin{equation}
C_{P} = T\left(\frac{\partial S}{\partial T}\right)_{P} = 2\pi r_{+}^{2} + \frac{4\pi}{3g^{2}r_{+}^{2}}(Q_{eff}^{2}+r_{+}^{2})\,.
\end{equation}
The heat flow at the upper isobar is given by:
\begin{equation}
Q_{H} = \int_{T_{1}}^{T_{2}} C_{P}{(P_{1},T)}dT = \int_{r_{+,1}}^{r_{+,2}} C_{P}{(g_{1},r_{+})}\frac{dT}{dr_{+}}dr_{+}\,,
\end{equation}
and integrates to:
\begin{equation}
Q_{H} = \frac{1}{2}g_{1}^{2}(r_{+,2}^{3}-r_{+,1}^{3}) + \frac{1}{2}(r_{+,2}-r_{+,1}) + 2Q_{eff}^{2}\left(\frac{1}{r_{+,2}}-\frac{1}{r_{+,1}}\right)\,.
\end{equation}
The work done during one cycle is simply the area of the rectangle inside the cycle:
\begin{equation}
W = (V_{2}-V_{1})(P_{1}-P_{4})\,.
\end{equation}
Explicitly,
\begin{equation}
W = \frac{1}{2}(g_{1}^{2}-g_{4}^{2})(r_{+,2}^{3}-r_{+,1}^{3}) + \frac{3}{2}Q_{eff}^{2}\left(1-\frac{g_{4}^{2}}{g_{1}^{2}}\right)\left(\frac{1}{r_{+,2}}-\frac{1}{r_{+,1}}\right)\,.
\end{equation}
The efficiency is then found to be:
\begin{equation}\label{etacycle2}
\eta = \frac{W}{Q_{H}} = \left(1-\frac{g_{4}^{2}}{g_{1}^{2}}\right)\left(1 - \frac{r_{+,2}-r_{+,1}}{g_{1}^{2}(r_{+,2}^{3}-r_{+,1}^{3})}\right)\,.
\end{equation}
As expected, the efficiency is again smaller than the Carnot efficiency. Moreover, the leading correction to the efficiency is charge-independent. If we convert $g$ and $r_{+}$ to $P$ and $S$, equation (\ref{etacycle2}) becomes
\begin{equation}
\eta = \left(1-\frac{P_{4}}{P_{1}}\right)\left(1-\frac{3}{8P_{1}}\left(\frac{S_{2}^{1/2}-S_{1}^{1/2}}{S_{2}^{3/2}-S_{1}^{3/2}}\right) \right)\,.
\end{equation}
This agrees with equation (20) in \cite{Johnson:2014yja} for the AdS-RN case.


\begin{thebibliography}{99}

\bibitem{Maldacena:1997re}
  J.~M.~Maldacena,
  ``The Large N limit of superconformal field theories and supergravity,''
  Int.\ J.\ Theor.\ Phys.\  {\bf 38}, 1113 (1999)
  [Adv.\ Theor.\ Math.\ Phys.\  {\bf 2}, 231 (1998)]
  [hep-th/9711200].

  \bibitem{Gubser:1998bc}
  S.~S.~Gubser, I.~R.~Klebanov and A.~M.~Polyakov,
  ``Gauge theory correlators from noncritical string theory,''
  Phys.\ Lett.\ B {\bf 428}, 105 (1998)
  [hep-th/9802109].

\bibitem{Witten:1998qj}
  E.~Witten,
  ``Anti-de Sitter space and holography,''
  Adv.\ Theor.\ Math.\ Phys.\  {\bf 2}, 253 (1998)
  [hep-th/9802150].

\bibitem{HawkingPage}
 S.~W.~Hawking and D.~N.~Page,
 ``Thermodynamics of Black Holes in anti-De Sitter Space,''
 Commun.\ Math.\ Phys.\ {\bf 87} 577 (1983).

  \bibitem{Witten:1998zw}
  E.~Witten,
  ``Anti-de Sitter space, thermal phase transition, and confinement in gauge theories,''
  Adv.\ Theor.\ Math.\ Phys.\  {\bf 2}, 505 (1998)
  [hep-th/9803131].

\bibitem{Chamblin:1999tk}
  A.~Chamblin, R.~Emparan, C.~V.~Johnson and R.~C.~Myers,
  ``Charged AdS black holes and catastrophic holography,''
  Phys.\ Rev.\ D {\bf 60}, 064018 (1999)
  [hep-th/9902170].

\bibitem{Cvetic:1999ne}
  M.~Cvetic and S.~S.~Gubser,
  ``Phases of R charged black holes, spinning branes and strongly coupled gauge theories,''
  JHEP {\bf 9904}, 024 (1999)
  [hep-th/9902195].

\bibitem{Kastor:2009wy}
  D.~Kastor, S.~Ray and J.~Traschen,
  ``Enthalpy and the Mechanics of AdS Black Holes,''
  Class.\ Quant.\ Grav.\  {\bf 26}, 195011 (2009)
  [arXiv:0904.2765 [hep-th]].

  \bibitem{Dolan:2011xt}
  B.~P.~Dolan,
  ``Pressure and volume in the first law of black hole thermodynamics,''
  Class.\ Quant.\ Grav.\  {\bf 28}, 235017 (2011)
  [arXiv:1106.6260 [gr-qc]].

  \bibitem{Dolan:2012jh}
  B.~P.~Dolan,
  ``Where is the PdV term in the fist law of black hole thermodynamics?,''
  arXiv:1209.1272 [gr-qc].

 \bibitem{Kubiznak:2012wp}
  D.~Kubiznak and R.~B.~Mann,
  ``P-V criticality of charged AdS black holes,''
  JHEP {\bf 1207}, 033 (2012)
  [arXiv:1205.0559 [hep-th]].

\bibitem{Hendi:2012um}
  S.~H.~Hendi and M.~H.~Vahidinia,
  ``Extended phase space thermodynamics and P-V criticality of black holes with a nonlinear source,''
  Phys.\ Rev.\ D {\bf 88}, no. 8, 084045 (2013)
  [arXiv:1212.6128 [hep-th]].

\bibitem{Kubiznak:2014zwa}
  D.~Kubiznak and R.~B.~Mann,
  ``Black Hole Chemistry,''
  arXiv:1404.2126 [gr-qc].

\bibitem{Xu:2014kwa}
  W.~Xu and L.~Zhao,
  ``Critical phenomena of static charged AdS black holes in conformal gravity,''
  Phys.\ Lett.\ B {\bf 736}, 214 (2014)
  [arXiv:1405.7665 [gr-qc]].

\bibitem{Grumiller:2014oha}
  D.~Grumiller, R.~McNees and J.~Salzer,
  ``Cosmological constant as confining U(1) charge in two-dimensional dilaton gravity,''
  Phys.\ Rev.\ D {\bf 90}, no. 4, 044032 (2014)
  [arXiv:1406.7007 [hep-th]].

\bibitem{Dolan:2014cja}
  B.~P.~Dolan,
  ``Bose condensation and branes,''
  JHEP {\bf 1410}, 179 (2014)
  [arXiv:1406.7267 [hep-th]].

\bibitem{Dolan:2014jva}
  B.~P.~Dolan,
  ``Black holes and Boyle's law — The thermodynamics of the cosmological constant,''
  Mod.\ Phys.\ Lett.\ A {\bf 30}, no. 03n04, 1540002 (2015)
  [arXiv:1408.4023 [gr-qc]].

  \bibitem{Rajagopal:2014ewa}
  A.~Rajagopal, D.~Kubiznak and R.~B.~Mann,
  ``Van der Waals black hole,''
  Phys.\ Lett.\ B {\bf 737}, 277 (2014)
  [arXiv:1408.1105 [gr-qc]].

\bibitem{Zhang:2014uoa}
  J.~L.~Zhang, R.~G.~Cai and H.~Yu,
  ``Phase transition and thermodynamical geometry for Schwarzschild AdS black hole in AdS$_{5}$ × S$^{5}$ spacetime,''
  JHEP {\bf 1502}, 143 (2015)
  [arXiv:1409.5305 [hep-th]].

\bibitem{Hendi:2014kha}
  S.~H.~Hendi, S.~Panahiyan and B.~E.~Panah,
  ``$P-V$ criticality and geometrothermodynamics of black holes with Born-Infeld type nonlinear electrodynamics,''
  arXiv:1410.0352 [gr-qc].

\bibitem{Zhao:2014fea}
  H.~H.~Zhao, L.~C.~Zhang, M.~S.~Ma and R.~Zhao,
  ``Phase Transition and Clapeyon Equation of Black Hole in Higher Dimensional AdS Spacetime,''
  arXiv:1411.3554 [hep-th].

\bibitem{Zhao:2014owa}
  H.~H.~Zhao, L.~C.~Zhang, M.~S.~Ma and R.~Zhao,
  ``Two phase equilibrium in charged topological dilaton AdS black hole,''
  arXiv:1411.7202 [gr-qc].

  \bibitem{Delsate:2014zma}
  T.~Delsate and R.~Mann,
  ``Van Der Waals Black Holes in $d$ dimensions,''
  JHEP {\bf 1502}, 070 (2015)
  [arXiv:1411.7850 [gr-qc]].

\bibitem{Zhang:2015ova}
  J.~L.~Zhang, R.~G.~Cai and H.~Yu,
  ``Phase transition and thermodynamical geometry of Reissner-Nordström-AdS black holes in extended phase space,''
  Phys.\ Rev.\ D {\bf 91}, no. 4, 044028 (2015)
  [arXiv:1502.01428 [hep-th]].

\bibitem{Hendi:2015kza}
  S.~H.~Hendi and Z.~Armanfard,
  ``Extended phase space thermodynamics and $P-V$ criticality of charged black holes in Brans-Dicke theory,''
  arXiv:1503.07070 [gr-qc].

 \bibitem{Johnson:2014xza}
  C.~V.~Johnson,
  ``Thermodynamic Volumes for AdS-Taub-NUT and AdS-Taub-Bolt,''
  Class.\ Quant.\ Grav.\  {\bf 31}, 235003 (2014)
  [arXiv:1405.5941 [hep-th]].

  \bibitem{Johnson:2014pwa}
  C.~V.~Johnson,
  ``The Extended Thermodynamic Phase Structure of Taub-NUT and Taub-Bolt,''
  Class.\ Quant.\ Grav.\ {\bf 31} 225005 (2014)
  [arXiv:1406.4533 [hep-th]].

\bibitem{Lee:2014tma}
  C.~O.~Lee,
  ``The extended thermodynamic properties of Taub–NUT/Bolt–AdS spaces,''
  Phys.\ Lett.\ B {\bf 738}, 294 (2014)
  [arXiv:1408.2073 [hep-th]].

\bibitem{Cai:2013qga}
  R.~G.~Cai, L.~M.~Cao, L.~Li and R.~Q.~Yang,
  ``P-V criticality in the extended phase space of Gauss-Bonnet black holes in AdS space,''
  JHEP {\bf 1309}, 005 (2013)
  [arXiv:1306.6233 [gr-qc]].

\bibitem{Frassino:2014pha}
  A.~M.~Frassino, D.~Kubiznak, R.~B.~Mann and F.~Simovic,
  ``Multiple Reentrant Phase Transitions and Triple Points in Lovelock Thermodynamics,''
  JHEP {\bf 1409}, 080 (2014)
  [arXiv:1406.7015 [hep-th]].

\bibitem{Dolan:2014vba}
  B.~P.~Dolan, A.~Kostouki, D.~Kubiznak and R.~B.~Mann,
  ``Isolated critical point from Lovelock gravity,''
  Class.\ Quant.\ Grav.\  {\bf 31}, no. 24, 242001 (2014)
  [arXiv:1407.4783 [hep-th]].

  \bibitem{Sherkatghanad:2014hda}
  Z.~Sherkatghanad, B.~Mirza, Z.~Mirzaeyan and S.~A.~H.~Mansoori,
  ``Critical behaviors and phase transitions of black holes in higher order gravities and extended phase spaces,''
  arXiv:1412.5028 [gr-qc].

\bibitem{Hendi:2015oqa}
  S.~H.~Hendi, S.~Panahiyan and M.~Momennia,
  ``Extended phase space of AdS Black Holes in Einstein-Gauss-Bonnet gravity with a quadratic nonlinear electrodynamics,''
  arXiv:1503.03340 [gr-qc].

\bibitem{Hennigar:2015esa}
  R.~A.~Hennigar, W.~G.~Brenna and R.~B.~Mann,
  ``$P − V$ criticality in quasitopological gravity,''
  JHEP {\bf 1507}, 077 (2015)
  [arXiv:1505.05517 [hep-th]].

\bibitem{Johnson:2014yja}
  C.~V.~Johnson,
  ``Holographic Heat Engines,''
  Class.\ Quant.\ Grav.\  {\bf 31}, 205002 (2014)
  [arXiv:1404.5982 [hep-th]].

\bibitem{Johnson:2013dka}
  C.~V.~Johnson,
  ``Large N Phase Transitions, Finite Volume, and Entanglement Entropy,''
  JHEP {\bf 1403}, 047 (2014)
  [arXiv:1306.4955 [hep-th]].

\bibitem{Aprile:2010ge}
  F.~Aprile, D.~Rodriguez-Gomez and J.~G.~Russo,
  ``p-wave Holographic Superconductors and five-dimensional gauged Supergravity,''
  JHEP {\bf 1101}, 056 (2011)
  [arXiv:1011.2172 [hep-th]].

\bibitem{Aprile:2011uq}
  F.~Aprile, D.~Roest and J.~G.~Russo,
  ``Holographic Superconductors from Gauged Supergravity,''
  JHEP {\bf 1106}, 040 (2011)
  [arXiv:1104.4473 [hep-th]].

\bibitem{Bobev:2011rv}
  N.~Bobev, A.~Kundu, K.~Pilch and N.~P.~Warner,
  ``Minimal Holographic Superconductors from Maximal Supergravity,''
  JHEP {\bf 1203}, 064 (2012)
  [arXiv:1110.3454 [hep-th]].

  \bibitem{Cvetic:2010jb}
  M.~Cvetic, G.~W.~Gibbons, D.~Kubiznak and C.~N.~Pope,
  ``Black Hole Enthalpy and an Entropy Inequality for the Thermodynamic Volume,''
  Phys.\ Rev.\ D {\bf 84}, 024037 (2011)
  [arXiv:1012.2888 [hep-th]].

\bibitem{Behrndt:1998jd}
  K.~Behrndt, M.~Cvetic and W.~A.~Sabra,
  ``Nonextreme black holes of five-dimensional N=2 AdS supergravity,''
  Nucl.\ Phys.\ B {\bf 553}, 317 (1999)
  [hep-th/9810227].

\bibitem{Duff:1999gh}
  M.~J.~Duff and J.~T.~Liu,
  ``Anti-de Sitter black holes in gauged N = 8 supergravity,''
  Nucl.\ Phys.\ B {\bf 554}, 237 (1999)
  [hep-th/9901149].

\bibitem{Cvetic:1999xp}
  M.~Cvetic, M.~J.~Duff, P.~Hoxha, J.~T.~Liu, H.~Lu, J.~X.~Lu, R.~Martinez-Acosta and C.~N.~Pope {\it et al.},
  ``Embedding AdS black holes in ten-dimensions and eleven-dimensions,''
  Nucl.\ Phys.\ B {\bf 558}, 96 (1999)
  [hep-th/9903214].

\bibitem{Gibbons:2005vp}
  G.~W.~Gibbons, M.~J.~Perry and C.~N.~Pope,
  ``Bulk/boundary thermodynamic equivalence, and the Bekenstein and cosmic-censorship bounds for rotating charged AdS black holes,''
  Phys.\ Rev.\ D {\bf 72}, 084028 (2005)
  [hep-th/0506233].

\bibitem{uvir1}
  L.~Susskind and E.~Witten,
  ``The Holographic Bound In Anti-De Sitter Space,''
  arXiv:hep-th/9805114.

\bibitem{uvir2}
  A.~W.~Peet and J.~Polchinski,
  ``UV/IR relations in AdS dynamics,''
  Phys.\ Rev.\ D {\bf 59} (1999) 065011
  [arXiv:hep-th/9809022].

\bibitem{uvir3}
  Y.~Hatta, E.~Iancu, A.~H.~Mueller and D.~N.~Triantafyllopoulos,
  ``Aspects of the UV/IR correspondence: energy broadening and string fluctuations,''
  JHEP {\bf 1102}, 065 (2011)
  [arXiv:1011.3763 [hep-th]].

\bibitem{uvir4}
  C.~A.~Ag\'on, A.~Guijosa and J.~F.~Pedraza,
  ``Radiation and a dynamical UV/IR connection in AdS/CFT,''
  JHEP {\bf 1406}, 043 (2014)
  [arXiv:1402.5961 [hep-th]].

\bibitem{progress}
 E.~Caceres, P.~H.~Nguyen and J.~F.~Pedraza,
 \emph{work in progress}.

  \bibitem{Ryu:2006bv}
  S.~Ryu and T.~Takayanagi,
  ``Holographic derivation of entanglement entropy from AdS/CFT,''
  Phys.\ Rev.\ Lett.\  {\bf 96}, 181602 (2006)
  [hep-th/0603001].

  \bibitem{Ryu:2006ef}
  S.~Ryu and T.~Takayanagi,
  ``Aspects of Holographic Entanglement Entropy,''
  JHEP {\bf 0608}, 045 (2006)
  [hep-th/0605073].

\bibitem{Lewkowycz:2013nqa}
  A.~Lewkowycz and J.~Maldacena,
  ``Generalized gravitational entropy,''
  JHEP {\bf 1308}, 090 (2013)
  [arXiv:1304.4926 [hep-th]].

  \bibitem{Hubeny:2007xt}
  V.~E.~Hubeny, M.~Rangamani and T.~Takayanagi,
  ``A Covariant holographic entanglement entropy proposal,''
  JHEP {\bf 0707}, 062 (2007)
  [arXiv:0705.0016 [hep-th]].

\bibitem{Haehl:2014zoa}
  F.~M.~Haehl, T.~Hartman, D.~Marolf, H.~Maxfield and M.~Rangamani,
  ``Topological aspects of generalized gravitational entropy,''
  JHEP {\bf 1505}, 023 (2015)
  [arXiv:1412.7561 [hep-th]].

  \bibitem{Hubeny:2013gta}
  V.~E.~Hubeny, H.~Maxfield, M.~Rangamani and E.~Tonni,
  ``Holographic entanglement plateaux,''
  JHEP {\bf 1308}, 092 (2013)
  [arXiv:1306.4004, arXiv:1306.4004 [hep-th]].

  \bibitem{toappear}
  E.~Caceres, J.~F.~Pedraza and J.~Virrueta,
 ``Finite volume effects in holographic entanglement entropy and mutual information,''
 \emph{to appear}.

\bibitem{Albash:2012pd}
  T.~Albash and C.~V.~Johnson,
  ``Holographic Studies of Entanglement Entropy in Superconductors,''
  JHEP {\bf 1205}, 079 (2012)
  [arXiv:1202.2605 [hep-th]].

\bibitem{Kol:2014nqa}
  U.~Kol, C.~Nunez, D.~Schofield, J.~Sonnenschein and M.~Warschawski,
  ``Confinement, Phase Transitions and non-Locality in the Entanglement Entropy,''
  JHEP {\bf 1406}, 005 (2014)
  [arXiv:1403.2721 [hep-th]].

\bibitem{Georgiou:2015pia}
  G.~Georgiou and D.~Zoakos,
  ``Entanglement entropy of the Klebanov-Strassler model with dynamical flavors,''
  JHEP {\bf 1507}, 003 (2015)
  [arXiv:1505.01453 [hep-th]].

\bibitem{MacDonald:2014zaa}
  S.~MacDonald,
  ``Thermodynamic Volume of Kerr-bolt-AdS Spacetime,''
  arXiv:1406.1257 [hep-th].

\bibitem{Brenna:2015pqa}
  W.~G.~Brenna, R.~B.~Mann and M.~Park,
  ``Mass and Thermodynamic Volume in Lifshitz Spacetimes,''
  arXiv:1505.06331 [hep-th].

\bibitem{Galante:2012pv}
  D.~Galante and M.~Schvellinger,
  ``Thermalization with a chemical potential from AdS spaces,''
  JHEP {\bf 1207}, 096 (2012)
  [arXiv:1205.1548 [hep-th]].

\bibitem{Caceres:2012em}
  E.~Caceres and A.~Kundu,
  ``Holographic Thermalization with Chemical Potential,''
  JHEP {\bf 1209}, 055 (2012)
  [arXiv:1205.2354 [hep-th]].

\bibitem{Caceres:2013dma}
  E.~Caceres, A.~Kundu, J.~F.~Pedraza and W.~Tangarife,
  ``Strong Subadditivity, Null Energy Condition and Charged Black Holes,''
  JHEP {\bf 1401}, 084 (2014)
  [arXiv:1304.3398 [hep-th]].

\bibitem{Caceres:2014pda}
  E.~Caceres, A.~Kundu, J.~F.~Pedraza and D.~L.~Yang,
  ``Weak Field Collapse in AdS: Introducing a Charge Density,''
  JHEP {\bf 1506}, 111 (2015)
  [arXiv:1411.1744 [hep-th]].

\bibitem{Bhattacharyya:2014oha}
  A.~Bhattacharyya, S.~Shajidul Haque and Á.~Véliz-Osorio,
  ``Renormalized Entanglement Entropy for BPS Black Branes,''
  Phys.\ Rev.\ D {\bf 91}, no. 4, 045026 (2015)
  [arXiv:1412.2568 [hep-th]].

\bibitem{Dolan:2010ha}
  B.~P.~Dolan,
  ``The cosmological constant and the black hole equation of state,''
  Class.\ Quant.\ Grav.\  {\bf 28}, 125020 (2011)
  [arXiv:1008.5023 [gr-qc]].

\bibitem{Sadeghi:2015ksa}
  J.~Sadeghi and K.~Jafarzade,
  ``Heat Engine of black holes,''
  arXiv:1504.07744 [hep-th].

\end{thebibliography}
\end{document}